\documentclass[10pt,conference]{IEEEtran} 

\usepackage{tikz}
\usepackage{amsmath}
\usepackage{todonotes}
\usepackage{filecontents}
\usepackage{svg}
\usepackage{listings}
\usepackage{algpseudocode}
\usepackage{multirow}
\usepackage{makecell}
\usepackage{xspace}
\usepackage{float}
\usepackage{caption}
\graphicspath{ {./imgs/} }
\svgpath{{../imgs/}}
\usepackage{enumitem}
\usepackage{textcomp}
\usepackage{amsthm}
\theoremstyle{definition}
\newtheorem{definition}{Definition}[section]
\usepackage[ruled,vlined,linesnumbered]{algorithm2e}
\usepackage{setspace}

\SetCommentSty{mycommfont}
\usepackage{url}
\usepackage{hyperref}
\usepackage{booktabs}
\usepackage{colortbl}
\usepackage{tikz}
\usetikzlibrary{fit,calc}
\usepackage[T1]{fontenc}
\usepackage{subfig}

\usepackage{amssymb}
\usepackage[normalem]{ulem}
\usepackage{pifont}
\usepackage{flushend}
\newcommand{\xmark}{\ding{55}}%

\hypersetup{
    colorlinks,
    linkcolor={red!50!black},
    citecolor={blue!50!black},
    urlcolor={blue!80!black}
}

\usepackage{xspace}

\newcommand{\eg}{\hbox{\emph{e.g.}}\xspace}
\newcommand{\ie}{\hbox{\emph{i.e.}}\xspace}

\newcommand{\etc}{\hbox{\emph{etc.}}\xspace}

\newcommand{\toolname}{\textsc{Sand}\xspace}
\newcommand{\para}[1]{\smallskip\noindent\textbf{#1}\xspace}
\newcommand{\asan}{ASan\xspace}
\newcommand{\ubsan}{UBSan\xspace}
\newcommand{\msan}{MSan\xspace}
\newcommand{\debloat}{Debloat\xspace}

\newcommand{\mpgain}{\texttt{mp3gain}\xspace}
\newcommand{\tiffsplit}{\texttt{{tiffsplit}\xspace}}
\newcommand{\mpaac}{\texttt{mp42aac}\xspace}
\newcommand{\infotocap}{\texttt{infotocap}\xspace}
\newcommand{\mujs}{\texttt{mujs}\xspace}
\newcommand{\pdftotext}{\texttt{pdftotext}\xspace}
\newcommand{\nm}{\texttt{nm}\xspace}
\newcommand{\objdump}{\texttt{objdump}\xspace}
\newcommand{\lame}{\texttt{lame}\xspace}
\newcommand{\exiv}{\texttt{exiv2}\xspace}

\newcommand{\ffmpeg}{\texttt{ffmpeg}\xspace}

\newcommand{\sqlite}{\texttt{sqlite3}\xspace}
\newcommand{\cflow}{\texttt{cflow}\xspace}
\newcommand{\jq}{\texttt{jq}\xspace}

\newcommand{\unifuzz}{\textsc{UniFuzz}\xspace}

\colorlet{tablecellcolor}{green!20} 
\newcommand{\green}[1]{\colorbox{green!20}{#1}}

\newcommand{\revise}[1]{#1}
\newcommand{\fuzzpost}{AFL++-POST\xspace}
\newcommand{\vdtest}{$\hat{A}_{12}$\xspace}
\newcommand{\camera}[1]{#1}

\usepackage[frozencache,cachedir=.]{minted}
\setminted{numbersep=5pt}
\usepackage[misc,geometry]{ifsym} 
\begin{document}

\title{\toolname: 
Decoupling Sanitization from Fuzzing \\for Low Overhead
}

\author{
\IEEEauthorblockN{Ziqiao Kong\IEEEauthorrefmark{2}\IEEEauthorrefmark{3}}
\IEEEauthorblockA{\small
ETH Zurich \text{and} \\
Nanyang Technological University \\
ziqiao001@e.ntu.edu.sg}
\and
\IEEEauthorblockN{Shaohua Li\IEEEauthorrefmark{2} \textsuperscript{\Letter}}
\IEEEauthorblockA{\small The Chinese University
of Hong Kong \\
shaohuali@cuhk.edu.hk}

\and
\IEEEauthorblockN{Heqing Huang}
\IEEEauthorblockA{\small City University of Hong Kong\\
heqhuang@cityu.edu.hk}

\and
\IEEEauthorblockN{Zhendong Su}
\IEEEauthorblockA{\small ETH Zurich\\
zhendong.su@inf.ethz.ch}
}

\maketitle
\begingroup\renewcommand\thefootnote{}
\footnotetext{\IEEEauthorrefmark{2} {Equal contribution}, \Letter\ Corresponding author}
\footnotetext{\IEEEauthorrefmark{3} {This work was done when Ziqiao Kong was a Master's student at ETH Zurich.}}

\begin{abstract}
Sanitizers provide robust test oracles for various vulnerabilities.
Fuzzing on sanitizer-enabled programs has been the best practice to find software bugs.
Since sanitizers require heavy program instrumentation to insert run-time checks, sanitizer-enabled programs have much higher overhead compared to normally built programs.

In this paper, we present \toolname, a new fuzzing framework that decouples sanitization from the fuzzing loop.
\toolname performs fuzzing on a \emph{normally built program} and only invokes the \emph{sanitizer-enabled program} when input is shown to be interesting.
Since most of the generated inputs are not interesting, \ie, not bug-triggering, \toolname allows most of the fuzzing time to be spent on the normally built program. We further introduce \emph{execution pattern} to practically and effectively identify interesting inputs.

We implement \toolname on top of AFL++ and evaluate it on 20 real-world programs.
Our extensive evaluation highlights its effectiveness: \revise{in 24 hours, compared to all the baseline fuzzers, \toolname significantly discovers more bugs while not missing any.}
\end{abstract}

\section{Introduction}

Fuzzing has been one of the most successful approaches to finding security vulnerabilities~\cite{clusterfuzz,zhu2022fuzzing}.
At a high level, fuzzers generate a large number of new inputs and execute the target program on each of them. Fuzzers typically rely on observable test oracles such as crashes to report bugs. 
However, many security flaws do not always yield crashes and thus are not detectable.
Sanitizers are designed to tackle this problem. When sanitizers are enabled at compile-time, compilers heavily instrument the target program to insert various checks. At run-time, violations of these checks result in program crashes.
Fuzzing on such \emph{sanitizer-enabled programs} is thus more effective in discovering software bugs.
To date, the most widely-used sanitizers include AddressSanitizer (\asan)~\cite{asan}, UndefinedBehaviorSanitizer (\ubsan)~\cite{ubsan}, and MemorySanitizer (\msan)~\cite{msan}. 

\medskip
\para{Problems.}
Sanitizers, despite their extraordinary bug-finding capability, have two main drawbacks. 
\textbf{\textit{First}}, sanitizers bring significant performance overhead to fuzzing. As our evaluation in Section~\ref{sec:san-overhead} will show, \asan, \ubsan, and \msan averagely slow down fuzzing speed by a factor of 3.3x, 2.0x, and 45x, respectively. 
Since fuzzing is computationally intensive, such high sanitizer overheads inevitably impede both the performance of fuzzers and the adoption of sanitizers.
Many approaches have been proposed to reduce the run-time overhead of sanitizers. For example, Debloat~\cite{debloat} optimizes \asan checks via sound static analysis. \textsc{SanRazor}~\cite{sanrazor} removes likely redundant \asan and \ubsan checks through dynamic profiling. FuZZan~\cite{fuzzan} designs dynamic metadata structure to improve the performance of \asan and \msan.
Notwithstanding these optimization efforts, the overhead imposed by sanitizers remains considerable. For instance, as our evaluation will show, the state-of-the-art effort, Debloat, can only reduce less than 10\% run-time cost of \asan.
Moreover, all these schemes require significant modifications to the existing sanitizer code base,
which hinders its compatibility with diverse infrastructures.
\textbf{\textit{Second}}, some sanitizers are mutually exclusive. For instance, because \asan and \msan maintain the same metadata structure, they can not be used together. Consequently, a fuzzer has to fuzz \asan-enabled programs and \msan-enabled programs separately to maximize its bug-finding capability.

\begin{figure}[tp]
\setminted[c]{highlightlines={1, 3, 17}, highlightcolor=gray!20}
    \begin{minted}[linenos,xleftmargin=1em,fontsize=\footnotesize,fontfamily=tt,escapeinside=@@]{c}
_TIFFfree(*read_ptr);
...
read_buff = *read_ptr;
if (!read_buff) {
    read_buff = limitMalloc(buffsize);
} else {
    if (prev_readsize < buffsize) {
        new_buff = _TIFFrealloc(read_buff, buffsize);
        if (!new_buff) {
            free(read_buff);
            read_buff = limitMalloc(buffsize);
        } else
            read_buff = new_buff;
    }
}

read_buff[buffsize] = 0;
    \end{minted}
\caption{A simplified Use-after-Free bug from libtiff in CVE-2023-26965. Line 17 triggers the bug because the freed buffer in line 1 is reallocated neither in line 5 nor in lines 8 and 11.}
\label{example:userafterfree}
\end{figure}

\smallskip
\para{Key insight.} 
Since sanitizers provide the security oracle, all current fuzzers execute sanitizer-enabled programs on every fuzzer-generated input to verify validity.
We now raise this question: \revise{\emph{Can we effectively filter bug-triggering inputs without executing a sanitizer-enabled program?}}
Theoretically, it seems paradoxical and infeasible as only by executing an input can we know if the input is bug-triggering.
However, our empirical evaluation will substantiate its feasibility.
\emph{The key insight is that bugs are strongly connected to execution paths.}
For instance, Figure~\ref{example:userafterfree} shows a simplified code snippet from CVE-2023-26965, which contains a Use-after-Free bug in line 17.
Normal and most execution paths are \{$1 \rightarrow 3 \rightarrow 5 \rightarrow 17$\}, \{$1 \rightarrow 3 \rightarrow 8 \rightarrow 10 \rightarrow 11 \rightarrow 17$\}, or \{$1 \rightarrow 3 \rightarrow 8 \rightarrow 13 \rightarrow 17$\}, where the freed buffer \verb|read_buff| in line 1 is correctly reallocated.
However, when the execution path is \{$1 \rightarrow 3 \rightarrow 17$\}, the freed buffer \verb|read_buff| is incorrectly used in line 17.
\revise{This buggy execution has a unique path not seen in other non-bug-triggering executions.}

Since triggering this bug requires exercising unique execution paths, our intuition is that we can encapsulate inputs with unique execution paths by executing them on normally built programs, then only feed these inputs with unique execution paths into sanitizer-enabled programs to reduce overall sanitization overhead.
More interestingly, our empirical evaluation in Section~\ref{sec:illustrative} will show that \emph{nearly all bugs can be accurately captured by unique execution paths.}
In fact, previous studies~\cite{igor,uaf} have also implicitly shown that bugs correlate highly to executions. 

\para{Our approach.}
Inspired by this observation, we propose a new fuzzing framework that decouples sanitization from the fuzzing loop for acceleration.
In the framework, the fuzzer (1) performs fuzzing on a binary that is built normally, \ie, without enabling sanitizers, then (2) selects inputs that have unique execution paths and runs them on the sanitizer-enabled binaries to check if they trigger any bugs.
Take the program shown in Figure~\ref{example:userafterfree} as an example: During fuzzing, when we first encounter an input that has the execution path \{$1 \rightarrow 3 \rightarrow 5 \rightarrow 17$\}, we re-execute the input on the sanitizer-enabled binary. The result is that this input does not trigger any bug. Thus, we will not validate all future inputs that have the same execution path on the sanitizer-enabled binary. When we first encounter an input that has the execution path \{$1 \rightarrow 3 \rightarrow 17$\}, similarly, we re-execute the input on the sanitizer-enabled binary. The result is that this input triggers a Use-after-Free bug.
As long as only (1) a small fraction of inputs have unique execution paths and (2) the buggy execution reliably has a unique execution path, we can significantly reduce the sanitization overhead during fuzzing.
As Section~\ref{sec:eval-pattern} will show, only 3.8\% of all inputs on average have unique execution paths, and more than 91\% of buggy executions have a unique execution path.

\smallskip
There is a key challenge in our approach: \emph{\textbf{How to efficiently obtain execution path during fuzzing?}}
Previous research has demonstrated that obtaining a fine-grained execution path is too costly to be practical~\cite{coverage_metrics, collafl}.
We tackle this problem by introducing \emph{execution pattern} to approximate \emph{execution path}. 
Execution pattern discards the order information in execution paths as a trade-off for efficiency. For instance, for the execution path \{$1 \rightarrow 3 \rightarrow 17$\}, its execution pattern is \{$1, 3, 17$\} meaning that code regions $1, 3$ and $17$ are executed.
Although such approximation may cause imprecision in theory, our evaluation will demonstrate that this design is accurate enough to identify unique execution paths.


\begin{table*}[h]
\begin{center}
\small
\renewcommand{\arraystretch}{1.2}
\rowcolors{1}{}{gray!15}
\setlength{\tabcolsep}{5.5pt}
\caption{Execution speed, \ie, number of executions per second, of native programs and sanitizer-enabled programs. The column ``Slowdown'' refers to the ratio of the native speed to the sanitizer speed. It is calculated by dividing the native speed by the sanitizer speed.
\xmark \ indicates a compilation failure. ``-'' indicates the incompatibility with the sanitizer.}
\vspace{-5pt}
\begin{tabular}{lr|rclrclrclrc}
\toprule[1.0pt]
    \multirow{2}{5em}{\textbf{Programs}} & \textbf{Native} & \multicolumn{2}{c}{\textbf{\asan}} && \multicolumn{2}{c}{\textbf{\debloat}} && \multicolumn{2}{c}{\textbf{\ubsan}} && \multicolumn{2}{c}{\textbf{\msan}}   \\
\cline{2-2}\cline{3-4}\cline{6-7}\cline{9-10}\cline{12-13}
    & \textbf{Speed} & \textbf{Speed}& \textbf{Slowdown} && \textbf{Speed} & \textbf{Slowdown} && \textbf{Speed} & \textbf{Slowdown} && \textbf{Speed} & \textbf{Slowdown} \\
\hline
cflow & 360 & 147 & 245\% && 164 & 220\%&& - & - && - & - \\
exiv2 & 298 & 92 & 324\% && \xmark & \xmark&& 225 & 133\% &&- & - \\
ffmpeg & 14 & 9 & 161\% && \xmark & \xmark&& 3 & 422\% &&- & - \\
gdk-pixbuf-pixdata & 237 & 68 & 348\% && 67 & 354\%&& 227 & 104\% && - & - \\
imginfo & 2,964 & 869 & 341\% && 907 & 327\%&& 1,968 & 151\% && 43 & 6,836\% \\
infotocap & 2,676 & 685 & 390\% && \xmark & \xmark&& 1,962 & 136\% &&43 & 6,209\% \\
jhead & 2,963 & 859 & 345\% && 888 & 334\%&& 2,652 & 112\% && 45 & 6,614\% \\
jq & 214 & 55 & 389\% && 53 & 401\%&& - & - && - & - \\
sqlite3 & 2,495 & 770 & 324\% && \xmark & \xmark&& 1,401 & 178\% &&- & - \\
lame & 101 & 64 & 157\% && 74 & 136\%&& - & - && - & - \\
mp3gain & 1,488 & 627 & 237\% && 634 & 235\%&& 917 & 162\% && 42 & 3,530\% \\
mp42aac & 1,917 & 472 & 406\% && \xmark & \xmark&& 682 & 281\% &&42 & 4,612\% \\
mujs & 1,491 & 425 & 351\% && 440 & 339\%&& 685 & 218\% && 42 & 3,590\% \\
nm & 2,209 & 586 & 377\% && \xmark & \xmark&& 1,597 & 138\% &&43 & 5,079\% \\
flvmeta & 4,100 & 1,356 & 302\% && 1,351 & 304\%&& 4,028 & 102\% && - & - \\
objdump & 573 & 212 & 270\% && \xmark & \xmark&& 250 & 229\% &&38 & 1,506\% \\
pdftotext & 410 & 151 & 271\% && \xmark & \xmark&& 192 & 214\% &&35 & 1,172\% \\
tcpdump & 1,754 & 432 & 407\% && 493 & 356\%&& 561 & 313\% && 42 & 4,196\% \\
tiffsplit & 2,093 & 665 & 315\% && \xmark & \xmark&& 1,247 & 168\% &&43 & 4,868\% \\
wav2swf & 2,757 & 486 & 604\% && 517 & 550\%&& 1,211 & 252\% && 43 & 6,357\% \\
\hline
\hiderowcolors
\textbf{Average} & \textbf{1,556} & \textbf{452} & \textbf{326\%} && \textbf{508} & \textbf{322\%} && \textbf{1,165} & \textbf{196\%} && \textbf{42} & \textbf{4,552\%} \\
\bottomrule
\vspace{-30px}
\end{tabular}

\label{table:san-overhead}
\end{center}
\end{table*}

Our idea is generally applicable to the gray-box fuzzer family.
Since AFL++~\cite{aflpp} is the most popular gray-box fuzzer in both academia and industry, we realized our idea on top of it and implemented a tool named \toolname.
We use 20 real-world programs widely used by the fuzzing community to evaluate \toolname. \revise{Our evaluation shows that in 24 hours, compared to all the baseline fuzzers, \toolname finds more bugs with statistical significance while not missing any bugs.}
In summary, we make the following contributions:
\begin{itemize}[leftmargin=15pt, topsep=5pt, itemsep=3pt]
    \item We identify that bugs are strongly connected with unique execution paths and further design an approximate yet accurate execution pattern to efficiently obtain execution path information during fuzzing.
    \item We propose a novel fuzzing framework that decouples sanitization from the fuzzing loop by selectively feeding fuzzer-generated inputs into sanitizers.
    \item We implement our idea in a tool named \toolname. We conduct in-depth evaluations to understand its effectiveness in terms of bug-finding, throughput, and coverage.
\end{itemize}

\section{Observation and Illustration}
In this section, we first show our observations on the high overhead of sanitizers and the rarity of bug-triggering inputs. Then, we use two real-world bug examples to illustrate the strong connections between bugs and execution paths.

\begin{table}[tp]
    \centering
    \footnotesize
    \caption{\revise{
    Ratio of bugger-triggering inputs. ``\#Exec.'' shows the number of all executions. ``\#Trig.'' shows the number of bug-triggering executions.
    }
    }
    \renewcommand{\arraystretch}{1.2}
    \setlength{\tabcolsep}{3pt}
{
\begin{tabular}[t]{lrrr}
        \toprule[1.0pt]
         \textbf{Programs} & \textbf{\#Exec.} & \textbf{\#Trig.} & \textbf{Ratio}  \\
         \hline
cflow & 13.2M & 1.2K& 0.0\% \\
exiv2 & 10.2M & 278 & 0.0\% \\
ffmpeg & 2.7M & 179 & 0.0\% \\
gdk-pixbuf. & 8.9M & 246K& 2.7\% \\
imginfo & 7.8M & 65K & 0.8\% \\
infotocap & 15.3M & 31K & 0.2\% \\
jhead & 16.1M & 404K & 2.5\% \\
jq & 4.7M & 425 & 0.0\% \\
sqlite3 & 17.4M & 350K & 2.1\% \\
lame & 4.4M & 875 & 0.0\% \\
\hline
    \end{tabular}
}\hfill
\begin{tabular}[t]{lrrr}
        \toprule[1.0pt]
         \textbf{Programs} & \textbf{\#Exec.} & \textbf{\#Trig.} & \textbf{Ratio}  \\
         \hline
mp3gain & 14.3M & 104K & 0.7\% \\
mp42aac & 3.6M & 13 & 0.0\% \\
mujs & 13.4M & 19K & 0.1\% \\
nm & 10.1M & 486 & 0.0\% \\
flvmeta & 23.8M & 531K & 2.2\% \\
objdump & 7.8M & 196K & 2.1\% \\
pdftotext & 5.8M & 1K & 0.0\% \\
tcpdump & 9.0M & 13K & 0.1\% \\
tiffsplit & 11.2M & 26K & 0.2\% \\
wav2swf & 8.3M & 451K & 6.3\% \\
\hline
\hiderowcolors
\textbf{Average} & \textbf{10.4M} & \textbf{122K} & \textbf{1.0\%} \\
\bottomrule
\vspace{-25px}
    \end{tabular}
\label{tab:bug-triggering}
\end{table}

\subsection{High Overhead of Sanitizers}\label{sec:san-overhead}

To benchmark sanitizer overhead in fuzzing, we use all 20 benchmark programs from our evaluation section. For each program, we compile five versions of it, \ie, native program, \asan-enabled program, \debloat-enabled program, \ubsan-enabled program, and \msan-enabled program. The native program refers to a normally built program without using any sanitizers. Since \debloat~\cite{debloat} achieves the state-of-the-art optimization for \asan, we include it to understand the significance of its improvement.
We use AFL++ as the default fuzzer. For each program, we:
\begin{enumerate}[label={\textbf{Step (\arabic*)}}, wide, labelwidth=!, labelindent=0pt, itemsep=5pt, topsep=5pt]
    \item Use AFL++ to fuzz the native program and collect the first \emph{one million} fuzzer-generated inputs. All these inputs are saved into disk. We use tmpfs~\cite{tmpfs} to reduce I/O overhead.
    \item Run AFL++ again on the native program to benchmark its running time on the saved one million inputs. AFL++ is adapted to fetch inputs from the disk instead of generation.
    \item Repeat \textbf{Step(2)} on four sanitizer-enabled programs to collect their running time on the same set of inputs.
\end{enumerate}

We ran the above experiment 10 times and reported the average fuzzing speed. All experimental settings are the same as our later evaluation in Section~\ref{sec:experiment-setup}.
Table~\ref{table:san-overhead} presents the average speed, \ie, number of executions per second, of each program.
Compared to native programs, \asan, \ubsan, and \msan averagely reduce the speed by 326\%, 196\%, and 4,552\%, respectively. 
Even for the best \asan optimization \debloat, its improvement over \asan is rather insignificant compared to the native programs.
Such huge sanitizer overheads inevitably hinder the fuzzing throughput. 

\subsection{Rareness of Bug-triggering Inputs}\label{sec:rare}
Fuzzers typically generate a large body of inputs for a target program.
It is intuitive that bug-triggering inputs are rarely met during fuzzing.
To understand the ratio of bug-triggering inputs to all the generated inputs, 
we count the number of all generated inputs as well as bug-triggering ones during 24 hours of fuzzing. The experimental data is from our later evaluation in Section~\ref{sec:bug-finding}.
Table~\ref{tab:bug-triggering} shows the result.
\revise{We can find that averagely \emph{only 1\%} of inputs are bug-triggering. For some programs, it is even rarer. For instance, on \pdftotext, less than 2 out of $10^4$ inputs trigger bugs.}
We can conclude that \textit{Only a tiny fraction of fuzzer-generated inputs are bug-triggering.}
Blindly sanitizing all of them is thus a huge waste.


\begin{figure}[tp]
\setminted[c]{highlightlines={8}, highlightcolor=gray!20}
    \begin{minted}[linenos,xleftmargin=1em,fontsize=\footnotesize,fontfamily=tt,escapeinside=@@]{c}
int wav_convert2mono(struct WAV *dest, int rate) 
{
    ...
    for(i=0; i < src->size; i += channels) {
      int j;
      int pos2 = ((int)pos)*2;
      for(j=0;j < fill; j += 2) {
        dest->data[pos2+j+0] = 0;
        dest->data[pos2+j+1] = src->data[i]+128;
      }
      pos += ratio;
    }
    ...
}
    \end{minted}
\caption{A simplified Buffer-Overflow bug from wav2swf in CVE-2017-11099. Line 8 triggers a buffer overflow when the \texttt{for} loops significantly change the buffer offset ``\texttt{pos2+j}''.}
\label{example:bufferoverflow}
\vspace{-15pt}
\end{figure}

\begin{figure}[tp]
\setminted[c]{highlightlines={9}, highlightcolor=gray!20}
    \begin{minted}[linenos,xleftmargin=1em,fontsize=\footnotesize,fontfamily=tt,escapeinside=@@]{c}
void JBIG2Stream::
readTextRegionSeg(Guint segNum, ...)
{
    ...
    numSyms = 0;
    for (i = 0; i < nRefSegs; ++i) {
      if ((seg = findSegment(refSegs[i]))) {
        if (seg->getType()==jbig2SegSymbol) {
           numSyms += seg->getSize();
        } else if (seg->getType() == jbig2Se) {
	      codeTables->append(seg);
        }
    ...
    syms = (JBIG2Bitmap **)gmallocn(numSyms);
    ...
}
    \end{minted}
\vspace{-5pt}
\caption{A simplified Integer-Overflow bug from xpdf (containing \pdftotext) in CVE-2022-38171. Line 9 triggers an integer overflow in \texttt{numSyms} when the \texttt{if} branch in line 8 is evaluated to \texttt{True} many times.}
\vspace{-10pt}
\label{example:integeroverflow}
\end{figure}

\subsection{Illustrative Examples}\label{sec:illustrative}
In Figure~\ref{example:userafterfree}, we have illustrated a Use-after-Free bug that has a unique execution path. In this section, we present motivating examples of different bug types.

\para{Buffer-Overflow.}
A Buffer-Overflow bug is triggered when buffer access exceeds the allocated range of stack or heap memory.
Figure~\ref{example:bufferoverflow} shows a real-world Buffer-Overflow bug from wav2swf in CVE-2017-11099.
The buggy buffer access is located in line 8. 
When the two \verb|for| loops iterate a significant number of times, the offset \verb|pos2+j| exceeds the buffer range of \verb|dest->data|.
The original cause is from the large values of both \verb|src->size| and \verb|fill|. But these unusual data subsequently lead to a unique execution path, \ie, a long chain of executions \{$1 \rightarrow 4 \rightarrow 7  \rightarrow 8 \rightarrow 9  \rightarrow 8 \rightarrow 9 \rightarrow \cdots \rightarrow 4 \rightarrow 7 \rightarrow 8 \rightarrow 9 \rightarrow \cdots$\}.
\emph{In practice, we observe that Buffer-Overflow bugs often accompany execution path changes that normal executions do not exercise.}
Thus, using unique execution paths as the indicator for sanitization can help us encapsulate bug-triggering inputs.
The many Buffer-Overflow bugs identified by \toolname in our evaluation will further confirm this rationale.

\para{Integer-Overflow.}
Figure~\ref{example:integeroverflow} shows an Integer-Overflow bug in line 9, where the variable \texttt{numSyms} overflows its valid range when the \texttt{if} guard in line 8 is frequently evaluated to true.
This bug leads to an unusual execution path. The overflowed value in \texttt{numSyms} subsequently causes a small allocated buffer in line 14, which leads to buffer overflow and \emph{dramatic path changes in the rest of the execution}.

\section{Our Approach}

This section introduces the design of our new fuzzing framework \toolname. Section~\ref{sec:executionpath} defines \emph{execution path} and its proxy approximation, \emph{execution pattern}. Section~\ref{sec:fuzzingframework} describes the fuzzing framework. Section~\ref{sec:implementation} clarifies technical details.

\subsection{Preliminary: Execution Path and its Proxy}\label{sec:executionpath}

Our approach is backed by the intuition that bug-triggering inputs have unique execution paths. 
We formally define the execution path as follows:



\begin{definition}[Execution Path]\label{def:execution-pattern}
Given an execution $\mathcal{E}$, the \emph{execution path} of $\mathcal{E}$ is defined as
${\Pi}_\mathcal{E} =\textbf{[}{e_1}, {e_2}, \cdots, {e_n}\textbf{]}$, where $e_i$ is the unique id of the code edge executed by $\mathcal{E}$. Note that, ${\Pi}_\mathcal{E}$ is ordered meaning that ${e_i}$ is executed before ${e_j}$ if $i<j$.
\end{definition}

\emph{Execution path} is a temporal transition sequence of all executed code when executing an input on the target program. It contains the full information on control-flow visits.
For instance, suppose a buggy execution \{$1 \rightarrow {2} \rightarrow 4  \rightarrow 3$\}. It has the execution path as \textbf{[}$1,2,4,3$\textbf{]}. Execution path is order-sensitive meaning that \textbf{[}$1,2,4,3$\textbf{]} $\neq$ \textbf{[}$1,4,2,3$\textbf{]}.
Unfortunately, obtaining the \emph{execution path} of execution is too expensive to be practical in fuzzing~\cite{coverage_metrics,collafl}. 
Since throughput is a key factor in fuzzing effectiveness, we cannot directly use \emph{execution path} in our design.
In this paper, we propose to use \emph{execution pattern} as an approximate yet accurate proxy for the execution path. We define \emph{execution pattern} as follows:

\begin{definition}[Execution Pattern]\label{def:execution-pattern}
Given an execution $\mathcal{E}$, the \emph{execution pattern} of $\mathcal{E}$ is defined as
$\mathcal{T}_\mathcal{E} =\{{e_1}, {e_2}, \cdots, {e_m}\}$, where ${e_i}\neq {e_j} (i \neq j)$ and $e_i$ is the unique id of the code edge reached by $\mathcal{E}$. Note that, $\mathcal{T}_\mathcal{E}$ is order-insensitive, \eg, $\{e_{1}, e_{2}, e_{3}\} = \{e_{2}, e_{3}, e_{1}\}$.
\end{definition}

\emph{Execution pattern} records all executed code edges of an execution. For example, the previous buggy execution \{$1 \rightarrow {2} \rightarrow 4 \rightarrow 3$\} has the execution pattern as
\{$1, 2, 4, 3$\}. Execution pattern is order-insensitive, \eg, \{$1, 2, 4, 3$\} = \{$1, 2, 3, 4$\}.
\revise{This execution pattern design can effectively approximate the execution path. Below, we discuss the soundness, realization, and alternative design of execution pattern in detail.}

\revise{
\smallskip
\noindent
\textbf{Soundness of execution pattern.}
Since \emph{execution pattern} abstracts over \emph{execution path}, it is theoretically possible that we cannot soundly capture all bug-triggering inputs with unique execution patterns.
Take Figure~\ref{example:bufferoverflow} as an example, where inputs taking the loop once or multiple times will have the same ``execution pattern''. However, this is only true within this specific function. 
The actual execution pattern considers the whole program, and consequently, the same local execution pattern does not represent the overall execution pattern. At a high level, such loop iteration differences result from or will result in different data (\eg, \texttt{src->size} and \texttt{pos}), which affect the execution of other parts of the program and eventually lead to divergent execution patterns. 
To further address this soundness concern, we provide an extensive evaluation in Section~\ref{sec:eval-pattern} to demonstrate that the proposed execution patterns can precisely filter bug-triggering inputs.}

\smallskip
\para{Realization of execution pattern.}
An essential benefit of the execution pattern is its ease of acquisition during fuzzing.
Fuzzers like AFL++ utilize an efficient data structure, \textit{bitmap}, to collect visited code edges of an execution. Figure~\ref{fig:bitmap} shows an example of this procedure. 
The bitmap is initialized to all zeros for a new execution. 
For the execution path [$5, 3, 9, 7$], the corresponding positions in the bitmap are marked to $1$. This bitmap is then used to update a global coverage map with a logic OR. 
For the next execution [$1, 7, 9, 2$], a similar bitmap is initialized and then marked. The coverage map is then cumulatively updated to record all code edges visited by all previous executions.
The design of \emph{execution pattern} allows us to obtain it effortlessly from the bitmap of execution, as shown in the middle right in Figure~\ref{fig:bitmap}.

\smallskip
\noindent
\revise{
\textbf{Alternative design of execution pattern.}
Theoretically, we can design different execution patterns to abstract execution paths. There are two key requirements to execution pattern design: (1) it should precisely distinguish bug-triggering inputs from normal inputs and (2) it is cheap to obtain during fuzzing.
Our current design meets all these requirements. 
Here, we discuss two possible alternatives.
The first one is \textbf{\emph{hit count}}, where we include the counts of visited code edges into execution patterns. In principle, this design can capture more bug-triggering inputs. However, as our evaluation in Section~\ref{sec:eval-pattern} will show, more than 13\% of the normal inputs also have a unique hit count, which means that \emph{hit count} does not meet the first requirement.
Another one is \textbf{\emph{coverage}}, where we simply use the code coverage as the execution pattern. An input has a unique execution pattern when it increases code coverage. Unfortunately, our evaluation in Section~\ref{sec:eval-pattern} shows that, on average, 55\% of bug-triggering inputs do not increase coverage, causing this design to miss bugs during fuzzing.
Our current design offers the first practical execution pattern implementation. Exploring a better design will improve the effectiveness of our approach and is orthogonal to our work.
}

\begin{figure}[tp]
    \centering
    \includegraphics[width=0.8\linewidth]{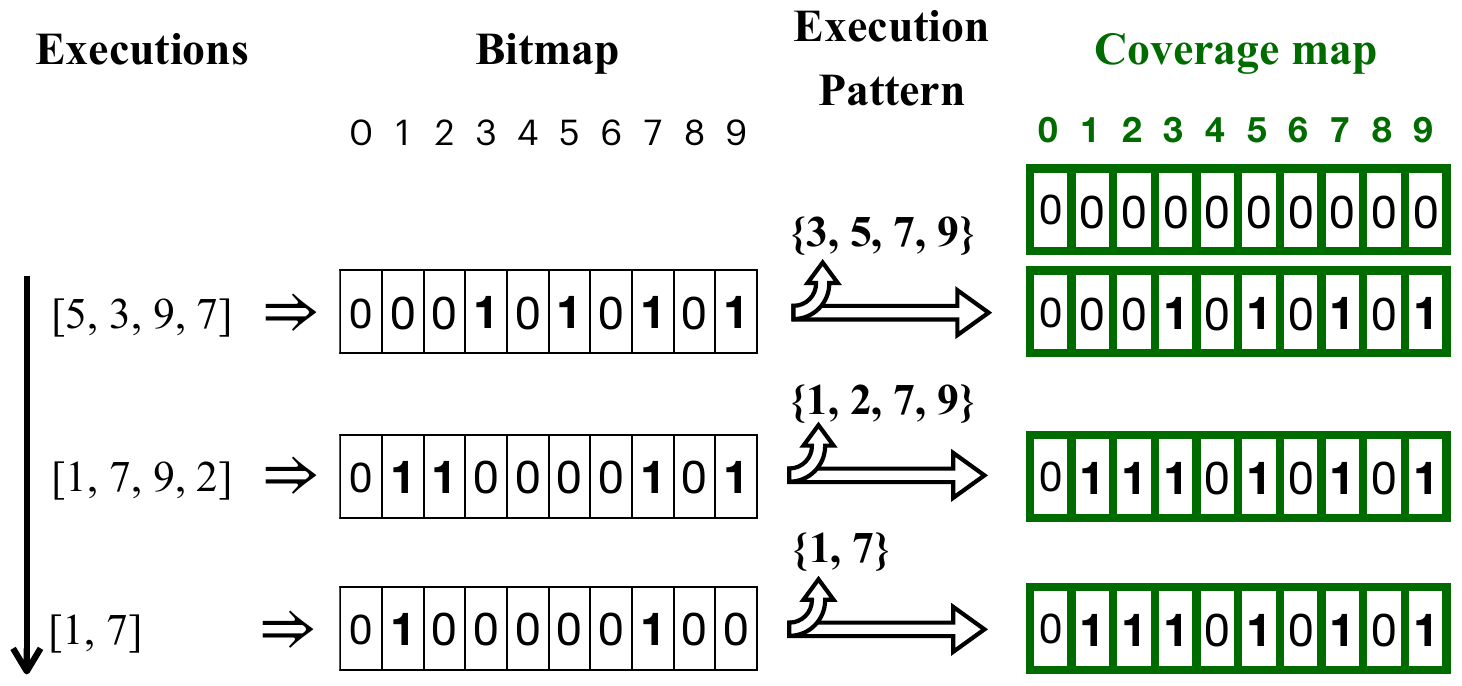}
    \caption{Executions (left) are recorded by bitmap (middle), which are used in AFL++ to update the coverage map (right). Our execution patterns can be derived from these bitmaps.}
    \label{fig:bitmap}
\end{figure}

\begin{figure}[tp]
    \centering
    \includegraphics[width=0.8\linewidth]{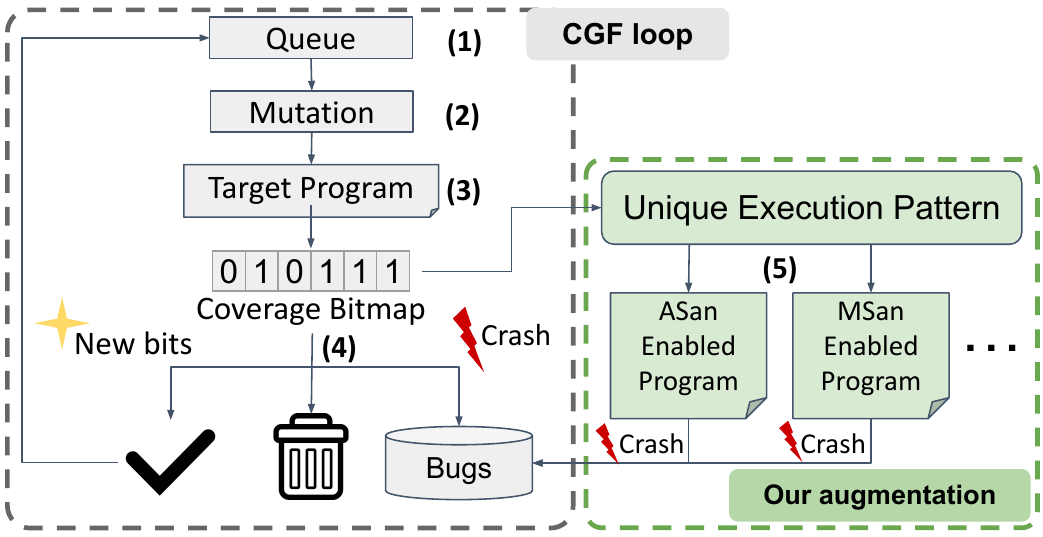}
    \caption{\toolname fuzzing loop.}
    \vspace{-15pt}
    \label{fig:workflow}
\end{figure}

\subsection{Sanitization-decoupled Fuzzing}\label{sec:fuzzingframework}

Based on the above formalization to execution patterns, we introduce our new fuzzing framework design.
We first describe the general workflow of a coverage-guided fuzzer (CGF). 
The gray area in Figure~\ref{fig:workflow} outlines the high-level sketch of a CGF. 
\emph{Before fuzzing starts, the fuzzer compiles the target program with fuzzer instrumentation and/or sanitizers.} 
Then, it fuzzes the target as follows:
\begin{enumerate}[label={\textbf{(\arabic*)}}, labelindent=0pt, itemsep=1pt, topsep=1pt]
    \item \textbf{Seed selection.} Select one seed from the seed pool according to predefined strategies.
    \item \textbf{Mutation.} Mutate the seed to generate new test inputs. 
    \item \textbf{Executing on the target program.} Execute a test input on the target program.
    \item \textbf{Coverage and execution analysis.}  Collect coverage feedback from the execution. If the execution increases coverage, save it to the seed pool; if the execution results in a crash, report the corresponding input as bug-triggering and save it to the disk; otherwise, discard it.
\end{enumerate}

As one can see, CGFs rely on the execution result of a target program to detect bugs. In order to maximize bug detection capability, current CGFs usually compile the target program with sanitizers enabled.
This routine significantly slows down fuzzing speed due to the high overhead of sanitizers.

In this paper, we tackle this problem by decoupling sanitization from the conventional fuzzing loop. 
The green part in Figure~\ref{fig:workflow} highlights our approach.
Before fuzzing starts, the fuzzer compiles multiple versions of the same program:
(1) a normally built program without any sanitizer enabled (denoted as $\mathcal{P}_{fuzz}$), \emph{on which the fuzzer performs fuzzing}, and (2) a set of sanitizer-enabled programs, \eg, \asan-enabled program ($\mathcal{P}_{\asan}$) and \msan-enabled program ($\mathcal{P}_{\msan}$).
The fuzzer follows the same steps as a CGF to fuzz the \emph{normally built program}.
But, after each execution of the target program, we introduce a new step:
\begin{enumerate}[label={\textbf{(\arabic*)}}, labelindent=3pt, itemsep=1pt, topsep=1pt]
\setcounter{enumi}{4}
\item \textbf{Conditional sanitization.} 
Extract the \emph{execution pattern} from the current execution's bitmap. If the execution pattern has been observed before, \ie, not unique, discard it. Otherwise, the current input is identified as \emph{sanitization-required}. The fuzzer then executes this input on each sanitizer-enabled program ($\mathcal{P}_{\asan}$, $\mathcal{P}_{\msan}$, \etc) and reports any discovered crashes.
\end{enumerate}

In conditional sanitization, we only consider the unique, \ie, first-seen, execution pattern to be sanitization-required.
One may be concerned that bug-triggering inputs do not always have a first-seen execution pattern, and thus, our design may potentially miss many bugs. 
To address this concern, we provide extensive experiments in Section~\ref{sec:evaluation} to demonstrate that (1) nearly all bug-triggering inputs (91.7\%) have unique execution patterns, and (2) our new fuzzing framework does not miss any bugs.
\camera{In practice, a single bug can be triggered hundreds or even thousands of times with unique different traces\cite{igor}, aligning with our observations during evaluation. Covering more than 91\% is sufficient to capture the bug at least one time, both empirically and evaluation proven. }

\para{Example.}
Figure~\ref{fig:uniqueexecutionpatterns} illustrates the process of identifying sanitization-required inputs. Starting from the first execution with pattern \{$3, 5, 7, 9$\}, the fuzzer identifies it as a unique execution pattern and thus sanitization-required. The second execution has the same pattern as before; thus, sanitization is unnecessary.
Similarly, the third and fifth executions have unique patterns not seen before and thus require sanitization.
Our hypothesis is that all inputs triggering unique bugs also have unique execution patterns.
Assuming that the fifth execution \{$6, 4, 2, 3, 5$\} is buggy, the fuzzer can successfully identify the bug during sanitization.
{\emph{Since executions holding the same execution path are likely to have similar semantics, \eg, exercising the same functionality or triggering the same bug,}} we only need to sanitize the execution with unique execution paths to identify the bug. 
This newly introduced \textbf{conditional sanitization} does not alter the standard fuzzing logic.
Execution patterns are obtained from the already-available bitmap collected on the normally built program. 

\colorlet{bg}{green!45}
\newcommand*{\tikzmk}[1]{\tikz[remember picture,overlay,] \node (#1) {};\ignorespaces}
\newcommand{\boxit}[3]{\tikz[remember picture,overlay]{

\node[xshift=5pt,yshift=3pt,fill=#1,opacity=0.20,fit={(A)($(B)+(.75\linewidth,.8\baselineskip)$)}] (box) {};

\node[text width=#3,anchor=south east, font=\footnotesize,align=right] at (box.south east) {#2}; 

}
\ignorespaces}

\begin{figure}[tp]
    \centering
    \includegraphics[width=0.7\linewidth]{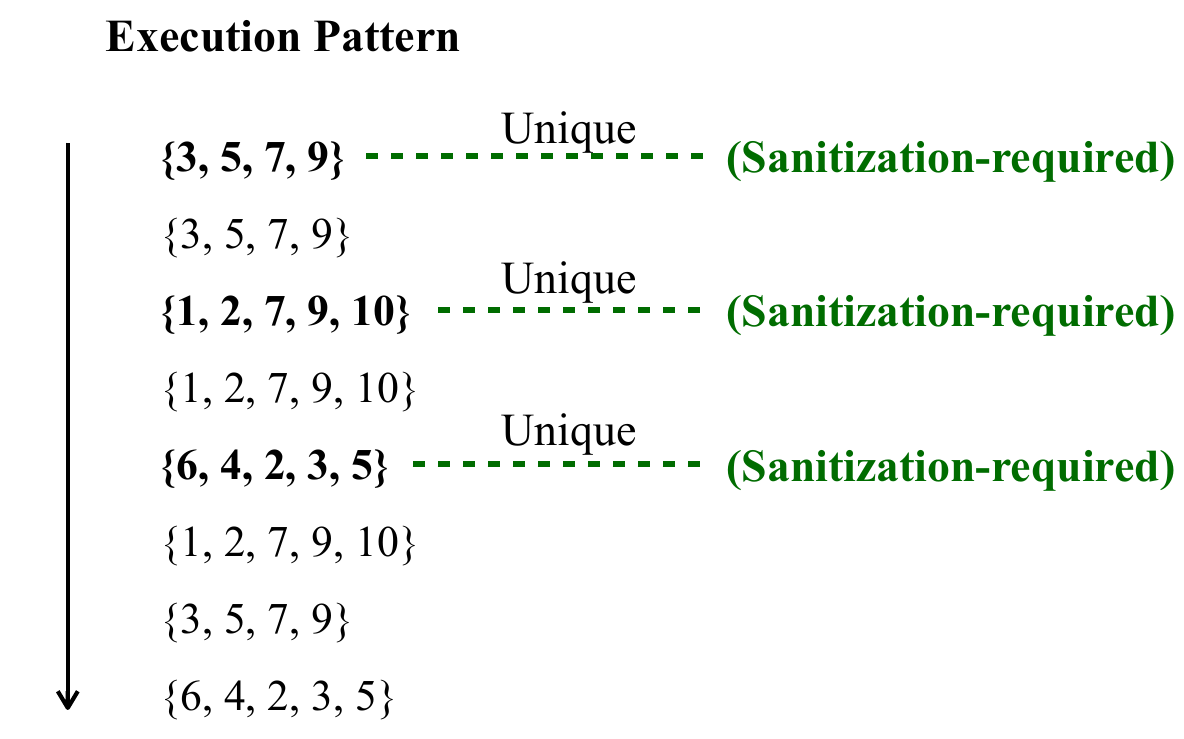}
    \vspace{-5pt}
    \caption{Out of eight consecutive executions from top to bottom, three are identified as unique and thus require sanitization.}
    \vspace{-5pt}
    \label{fig:uniqueexecutionpatterns}
    \vspace{-5pt}
\end{figure}

Algorithm \ref{alg:ourtool} sketches the implementation pseudo-code of our new fuzzing framework.
In each fuzzing loop (line 1), the fuzzer first selects a seed $s$ and mutates it to generate a new input $s'$ (lines 2-3). Next, it executes the normally built program $\mathcal{P}_{fuzz}$ on the input to collect its execution return $ret$ and $bitmap$ (line 4).
Then, it extracts the execution pattern $\mathcal{T}_{\mathcal{E}}$ from $bitmap$ (line 6) and determines whether or not this execution pattern has been observed (line 7). 
If $\mathcal{T}_{\mathcal{E}}$ is new, the fuzzer labels it as sanitization-required and executes each of the available sanitizer-enabled programs $\mathcal{P}_{san}$ on the input $s'$ (lines 8-9). Meanwhile, $\mathcal{T}_{\mathcal{E}}$ will be added to the hash table.
If any execution crashes, meaning the input $s'$ triggers a bug, the fuzzer sets the return status to \emph{crash} (lines 10-11).
Finally, the fuzzer continues the original procedure: save the new input as bug-triggering if the return status is \emph{crash} (lines 13-14); or queue it to the seed pool if it increases coverage (lines 15-16).

Our new fuzzing framework decouples sanitization from standard fuzzing logic. It has the following main advantages:
\begin{itemize}[leftmargin=10pt, topsep=5pt, itemsep=3pt]
    \item \textbf{\emph{Orthogonal to CGFs.}}
    We introduce only an additional step to execute sanitizer-enabled programs on selected inputs. Conceptually, our approach can augment any AFL-family fuzzers without modifying their main fuzzing logic.
    \item \textbf{\emph{Sanitizer inclusive.}}
    Some sanitizers like \asan and \msan are mutually exclusive, meaning that they cannot be used together on a program. Current fuzzers can only perform fuzzing on a program with only one of such sanitizers enabled. In our \toolname, multiple sanitizer-enabled programs can be used for sanitization simultaneously. 
    We will provide additional technical details in Section~\ref{sec:implementation} to explain how we support multiple sanitizers.
\end{itemize}




\begin{algorithm}[tp]
\setstretch{1.0}
\small
\SetFuncSty{sffamily}

\DontPrintSemicolon
\SetKwInput{KwInput}{Input}                
\SetKwInput{KwOutput}{Output}              
\SetKwFunction{Abort}{Abort}
\SetKwFunction{SelectSeed}{SelectSeed}
\SetKwFunction{Mutate}{Mutate}
\SetKwFunction{Execution}{Execute}
\SetKwFunction{SanExecution}{SanExecute}
\SetKwFunction{GetExecutionPattern}{GetExecutionPattern}
\SetKwFunction{IsUnique}{IsUnique}

\SetKwData{Prefix}{Pre}
\SetKwData{interesting}{interesting}

\KwInput{Seed pool $\mathcal{S}$.}

\While{$\neg\Abort()$}{
    $s \gets \SelectSeed(\mathcal{S})$ \tcp{\small Seed selection}
    $s' \gets \Mutate(s)$ \tcp{\small Generate input}
    $ret$, $bitmap$ $ \gets \Execution(s', \mathcal{P}_{fuzz})$\;
\;
    \tikzmk{A}$\mathcal{T}_{\mathcal{E}} \gets \GetExecutionPattern(bitmap)$\;
    \If{\IsUnique($\mathcal{T}_{\mathcal{E}}$)}
    {
        \ForEach{$\mathcal{P}_{san} \in \{\mathcal{P}_{\asan}, \mathcal{P}_{\msan}, \cdots\}$}{
            $ret_{san} \gets \SanExecution(s', \mathcal{P}_{san})$\;
            \If{$ret_{san} == crash$}{
                $ret = crash$\;
            }
        }
    }\tikzmk{B}\boxit{bg}{// Our augmentation}{3cm}

    \If(\tcp*[h]{\small Crash?}){$ret == crash$} 
    {
        save $s'$ to disk\;
    }
    \If(\tcp*[h]{\small~New coverage?}){\textup{covers new code}}{
        add $s'$ to $\mathcal{S}$\;
    }
}
\caption{The New Fuzzing Loop of \toolname}
\label{alg:ourtool}
\end{algorithm}

\begin{algorithm}[tp]
\setstretch{1.0}
\small
\SetFuncSty{textbf}
\SetDataSty{sffamily}

\DontPrintSemicolon

\SetKwProg{Fn}{}{:}{}
\SetKwFunction{IsUnique}{IsUnique}

\SetKwData{HashTable}{HashTable}
\SetKwData{Hash}{Hash}

\Fn{\IsUnique{$\mathcal{T}_{\mathcal{E}}$}}
{
    $cksum$ $\gets Hash(\mathcal{T}_{\mathcal{E}})$\;
    \If{$\HashTable[cksum] \neq 1$} 
    {
        $\HashTable[cksum] = 1$\;
        \Return $True$;
    }
    \Return $False$;
}
\caption{Identify unique execution patterns}
\label{alg:hash}
\end{algorithm}

\subsection{Implementation}\label{sec:implementation}
\noindent\textbf{Unique execution pattern analysis.}
We obtain the execution pattern of an execution from its bitmap. In our implementation, we use the \texttt{simplify\_trace()} function in AFL++ to achieve this goal.
This design allows us to efficiently get execution patterns during fuzzing.
To identify unique execution patterns, we calculate checksums of all observed execution patterns and use a hash map to store them. Algorithm~\ref{alg:hash} shows the pseudocode. 
The hash table {\sffamily{HashTable}} is initialized to all zeros at the start of fuzzing.
In our implementation, we use XXH32 hashing algorithm~\cite{xxhash} because of its fast speed. 
\revise{
Note that the hash function is used to identify unique execution patterns. 
In theory, any method that can separate unique patterns, such as bit-wise match, can be used here as an alternative solution to the hash function. 
However, the selected method will not affect the effectiveness of \toolname as long as (1) it has negligible overhead and (2) all unique execution patterns can be precisely captured. 
Our evaluation in Section~\ref{sec:hash-overhead} will show that the selected hash function meets all requirements.}

\smallskip
\noindent\textbf{Program instrumentation in \toolname.}
The fuzz target $\mathcal{P}_{fuzz}$ is instrumented by \toolname to include the necessary instrumentation code for coverage collection. 
Since all the sanitizer-enabled programs are used for sanitization only, no such instrumentation is needed. Thus, we directly use the LLVM compiler to compile  $\mathcal{P}_{\asan}$, $\mathcal{P}_{\ubsan}$, and $\mathcal{P}_{\msan}$.
Because \asan and \ubsan are compatible, we combine them as $\mathcal{P}_{\asan/\ubsan}$. 
To reduce the burden of invoking these programs, we utilize the \emph{forkserver}~\cite{zalewski2014afl} to create one forkserver to communicate with all sanitizer-enabled programs efficiently during fuzzing.

\begin{figure*}[!t]
    \centering
    \includegraphics[trim={0 30 0 40}, width=\linewidth, page=1]{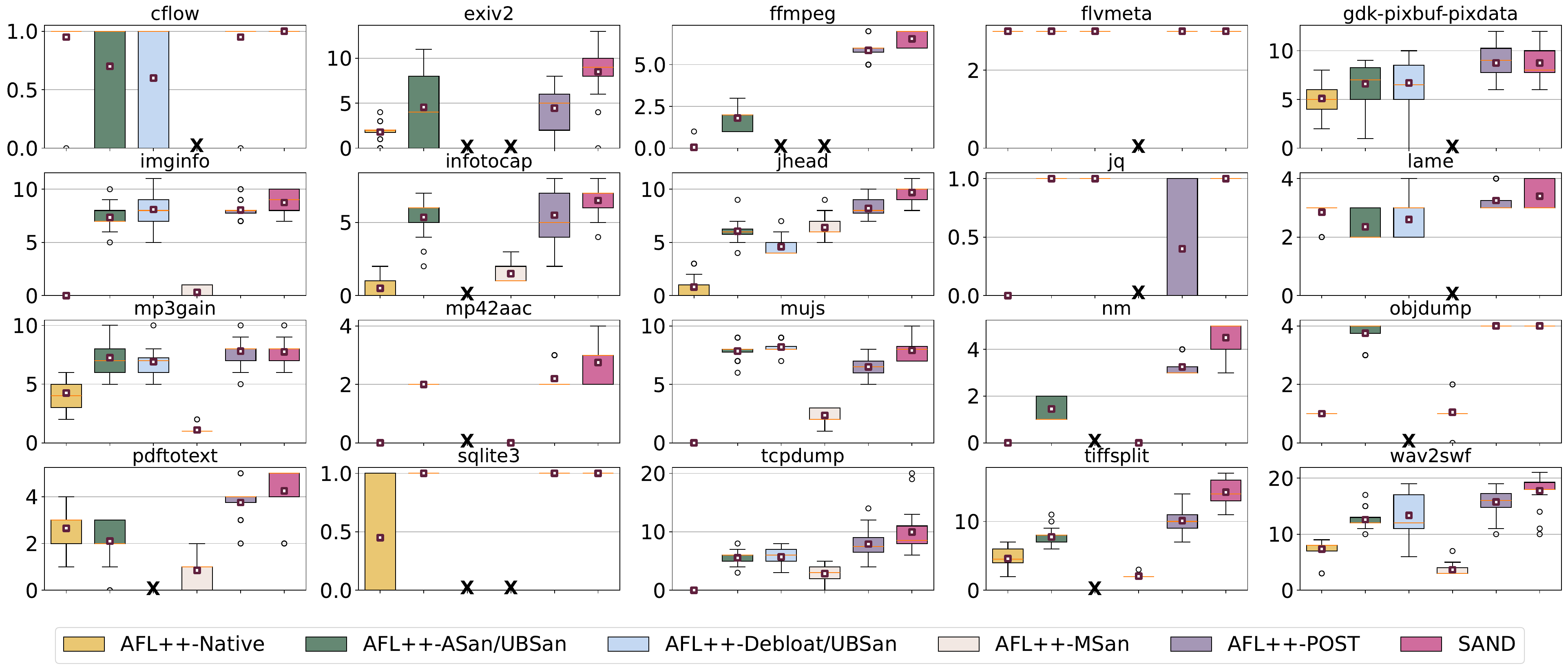}
    \caption{
    \revise{
    Number of unique detected bugs across repetitions. \xmark indicates a compilation failure or sanitizer incompatibility.}}
    \vspace{-10pt}
    \label{fig:box-bugs}
\end{figure*}

\section{Evaluation}\label{sec:evaluation}

 
We implemented \toolname based on AFL++-4.05c~\cite{aflpp}, the latest version at the time of implementation. AFL++ is the state-of-the-art gray-box fuzzer and has been widely used as the baseline fuzzer in many previous study~\cite{fuzzconcolic, pmfuzz, fuzzdp}. 

\subsection{Experimental Setup}\label{sec:experiment-setup}






\noindent\textbf{Benchmark.} 
We use real-world programs from the benchmarking test platform \unifuzz~\cite{unifuzz} for our evaluation. 
We use the provided seeds from \unifuzz for all fuzzing campaigns. 
To maximally understand \toolname's capability in different sanitizers, we use all three popular sanitizers, \ie, \asan, \ubsan, and \msan.
We evaluate all 20 programs on \asan and \ubsan.
Due to the compatibility issue of \msan, we failed to instrument 8 programs with \msan. We thus exclude \msan on their evaluation. For \cflow, \lame, and \jq, all the \unifuzz provided seeds will crash \ubsan due to unaligned pointers near program entry points, we thus use \asan instead of \asan/\ubsan for these three programs.
These programs cover a diverse range of input types, including
\begin{itemize}[leftmargin=15pt]
    \item \textbf{Image}: imginfo, jhead, tiffsplit, exiv2, gdk-pixbuf-pixdata.
    \item \textbf{Audio}: mp3gain, wav2swf, lame.
    \item \textbf{Video}: mp42aac, ffmpeg, flvmeta.
    \item \textbf{Text}: infotocap, mujs, pdftotext, cflow, jq, sqlite3.
    \item \textbf{Binary}: nm, objdump.
    \item \textbf{Network}: tcpdump.
\end{itemize}
\revise{
There are other available benchmarks, such as Magma~\cite{magma} and Fuzzbench~\cite{fuzzbench}.
We did not use them because (1) many sanitizer-reported bugs are not covered by Magma’s bug set since Magma uses manually analyzed bugs, (2) Magma contains only 9 projects while \unifuzz contains 20 projects, and (3) Fuzzbench’s programs contain too few bugs as its evaluation metric focuses on coverage. 
}



\para{Baseline.} 
Since \asan and \ubsan are compatible with each other, we combine them together when building binaries.
All fuzzers and programs are built with LLVM-14, the latest stable version at the time of implementation.
We compare the performance of \toolname against four baseline fuzzers:
\begin{enumerate}[leftmargin=15pt, itemsep=5pt]
    \item \textbf{AFL++-Native}: Fuzzing normally built programs 
    
    \item \textbf{AFL++-\asan/\ubsan}: Fuzzing $\mathcal{P}_{\asan/\ubsan}$.
    
    \item \textbf{AFL++-\msan}: Fuzzing $\mathcal{P}_{\msan}$.
    
    \item \textbf{AFL++-\debloat/\ubsan}: 
To understand if \toolname can surpass the existing sanitizer optimization schemes, we also choose the state-of-the-art \asan optimization technique, \debloat~\cite{debloat}. 
Because \debloat optimizes \asan, it can also be used together with \ubsan. To maximize its bug detection capability, we let AFL++ to fuzz on \debloat/\ubsan-enabled program (denoted as ``AFL++-\debloat/\ubsan''). All programs instrumented with \debloat are built with LLVM-12 because this is the highest LLVM version that \debloat supports. Since compiling \exiv, \ffmpeg, \infotocap, \mpaac, \sqlite, \nm, \objdump, \pdftotext and \tiffsplit with \debloat results in compilation or instrumentation failures, we exclude them for AFL++-\debloat/\ubsan.

    \item \revise{
    \textbf{\fuzzpost}: Another workaround to reduce sanitizer overhead is post-processing all the inputs saved in the corpus when fuzzing with AFL++-Native. These inputs increase coverage during fuzzing normally built programs. This baseline can also be viewed as using coverage-increasing information as the execution pattern. 
    }
\end{enumerate}

\para{Hardware and Setup.} We conduct all experiments on a machine equipped with an AMD 3990x CPU and 256G memory running Ubuntu 22.04. Following Klee's~\cite{klee} standard, \revise{we repeated all experiments 20 times} and ran all fuzzing campaigns for 24 hours. 

\begin{table}[!tp]
    \centering
    \small
    \setlength\tabcolsep{2pt}
    \renewcommand{\arraystretch}{1.0}
    \rowcolors{3}{}{gray!15}
    \captionof{table}{\revise{
    Mean number of unique bugs across repetitions. \xmark~ indicates a compilation failure, and ``-'' indicates incompatibility. The largest mean numbers are highlighted in \green{green}.}}
    \begin{tabular}{lrrrrrr}
        \toprule[1.0pt]
        \multirow{2}{*}[-0.5em]{\textbf{Programs}} & \multicolumn{5}{c}{\textbf{AFL++-}} & \multirow{2}{*}[-0.5em]{\textbf{\toolname}\textsubscript{$p$-val, \vdtest}} \\
        \cline{2-6}
        & \textbf{Native} & \makecell{\textbf{\asan/} \\ \textbf{\ubsan}} & \makecell{\textbf{Debloat/} \\ \textbf{\ubsan}} & \textbf{\msan} & \textbf{POST}&  \\
         \hline
cflow & 0.95 & 0.70 & 0.60 & \xmark & 0.95 & $ \green{1.00}_{0.34, 0.53} $ \\
exiv2 & 1.80 & 4.55 & \xmark & \xmark & 4.45 & $ \green{8.50}_{0.00, 0.89} $ \\
ffmpeg & 0.05 & 1.80 & \xmark & \xmark & 5.85 & $ \green{6.55}_{0.00, 0.78} $ \\
gdk. & 5.10 & 6.60 & 6.70 & \xmark & \green{8.75} & $ \green{8.75}_{0.92, 0.49} $ \\
imginfo & 0.00 & 7.35 & 8.10 & 0.30 & 8.05 & $ \green{8.75}_{0.04, 0.68} $ \\
infotocap & 0.50 & 5.35 & \xmark & 1.50 & 5.50 & $ \green{6.50}_{0.04, 0.68} $ \\
jhead & 0.80 & 6.05 & 4.60 & 6.40 & 8.20 & $ \green{9.70}_{0.00, 0.86} $ \\
jq & 0.00 & \green{1.00} & \green{1.00} & \xmark & 0.40 & $ \green{1.00}_{0.00, 0.80} $ \\
sqlite3 & 0.45 & \green{1.00} & \xmark & \xmark & \green{1.00} & $ \green{1.00}_{1.00, 0.50} $ \\
lame & 2.85 & 2.35 & 2.60 & \xmark & 3.25 & $ \green{3.40}_{0.33, 0.57} $ \\
mp3gain & 4.25 & 7.25 & 6.90 & 1.10 & \green{7.80} & $ 7.75_{0.72, 0.47} $ \\
mp42aac & 0.00 & 2.00 & \xmark & 0.00 & 2.20 & $ \green{2.75}_{0.01, 0.69} $ \\
mujs & 0.00 & 7.85 & \green{8.20} & 2.35 & 6.50 & $ 7.90_{0.00, 0.84} $ \\
nm & 0.00 & 1.45 & \xmark & 0.00 & 3.25 & $ \green{4.50}_{0.00, 0.90} $ \\
flvmeta & \green{3.00} & \green{3.00} & \green{3.00} & \xmark & \green{3.00} & $ \green{3.00}_{1.00, 0.50} $ \\
objdump & 1.00 & 3.75 & \xmark & 1.05 & \green{4.00} & $ \green{4.00}_{1.00, 0.50} $ \\
pdftotext & 2.65 & 2.10 & \xmark & 0.85 & 3.75 & $ \green{4.25}_{0.02, 0.70} $ \\
tcpdump & 0.00 & 5.55 & 5.70 & 2.85 & 7.90 & $ \green{9.95}_{0.08, 0.66} $ \\
tiffsplit & 4.60 & 7.75 & \xmark & 2.05 & 10.10 & $ \green{14.25}_{0.00, 0.97} $ \\
wav2swf & 7.35 & 12.60 & 13.35 & 3.65 & 15.75 & $ \green{17.75}_{0.00, 0.79} $ \\
\bottomrule
    \end{tabular}
    \label{tab:bugs-mean}
\vspace{-10pt}

\end{table}

\subsection{Bug-Finding Capability}\label{sec:bug-finding}








Finding bugs is the ultimate goal of fuzzing. In this section, we evaluate the bug-finding capability of all fuzzers. In particular, we would like to answer the following two questions:

\begin{enumerate}[label=\textbf{Q\arabic*}, leftmargin=20pt, topsep=5pt, itemsep=5pt]
    \item\label{rq:bug-num} Does \toolname find \emph{more bugs} compared to other fuzzers?
    \item\label{rq:bug-miss} Does \toolname \emph{miss any bugs} found by other fuzzers?
\end{enumerate}

To answer these questions, we collect all crashes found by each fuzzer.
We triage all crashes according to their root causes to quantify the number of unique bugs each fuzzer finds. 
Our deduplication is done via both the stack frame information from GDB~\cite{unifuzz} and manual analysis.

\smallskip
\para{New bugs.}
Before discussing the evaluation result of \toolname, we would like to mention that during the evaluation, we found nine new bugs using \toolname. All of these bugs have been confirmed by the developers. Three of them were assigned CVE numbers. Considering all the programs have been heavily fuzzed in both academia and industry, these new bugs reflect the effectiveness of our approach. 

\begin{figure*}[tp]
\begin{minipage}{1.0\linewidth}
\hfill
    \begin{minipage}{0.49\linewidth}
        \centering
        \small
        \setlength\tabcolsep{4pt}
        \renewcommand{\arraystretch}{1.1}
        \captionof{table}{\revise{
        Number of unique bugs found by each fuzzer. ``Total'' row shows the total number of unique bugs. ``$\toolname-$'' means the number of bugs missed by \toolname. ``$\toolname+$'' means the number of bugs additionally covered by \toolname.}}
        \begin{tabular}{lrrrrrr}
            \toprule[1.0pt]
            \multirow{2}{*}[-0.5em]{} & \multicolumn{5}{c}{\textbf{AFL++-}} & \multirow{2}{*}[-0.5em]{\textbf{\toolname}} \\
            \cline{2-6}
            & \textbf{Native} & \makecell{\textbf{\asan/} \\ \textbf{\ubsan}} & \makecell{\textbf{Debloat/} \\ \textbf{\ubsan}} & \textbf{\msan} & \textbf{POST}&  \\
             \hline
             \textbf{Total} & 70 & 145 & 100 & 54 & 174 & \textbf{204} \\
             \textbf{$\toolname-$} & 0 & 0 & 0 & 0 & 0 & - \\
             \textbf{$\toolname+$} & 134 & 59 & 104 & 150 & 30 & - \\
            \bottomrule
        \end{tabular}
        \label{fig:bugs-venn}
    \end{minipage}
\hfill
    \begin{minipage}{0.5\linewidth}
        \centering        \includegraphics[width=0.8\linewidth, page=1]{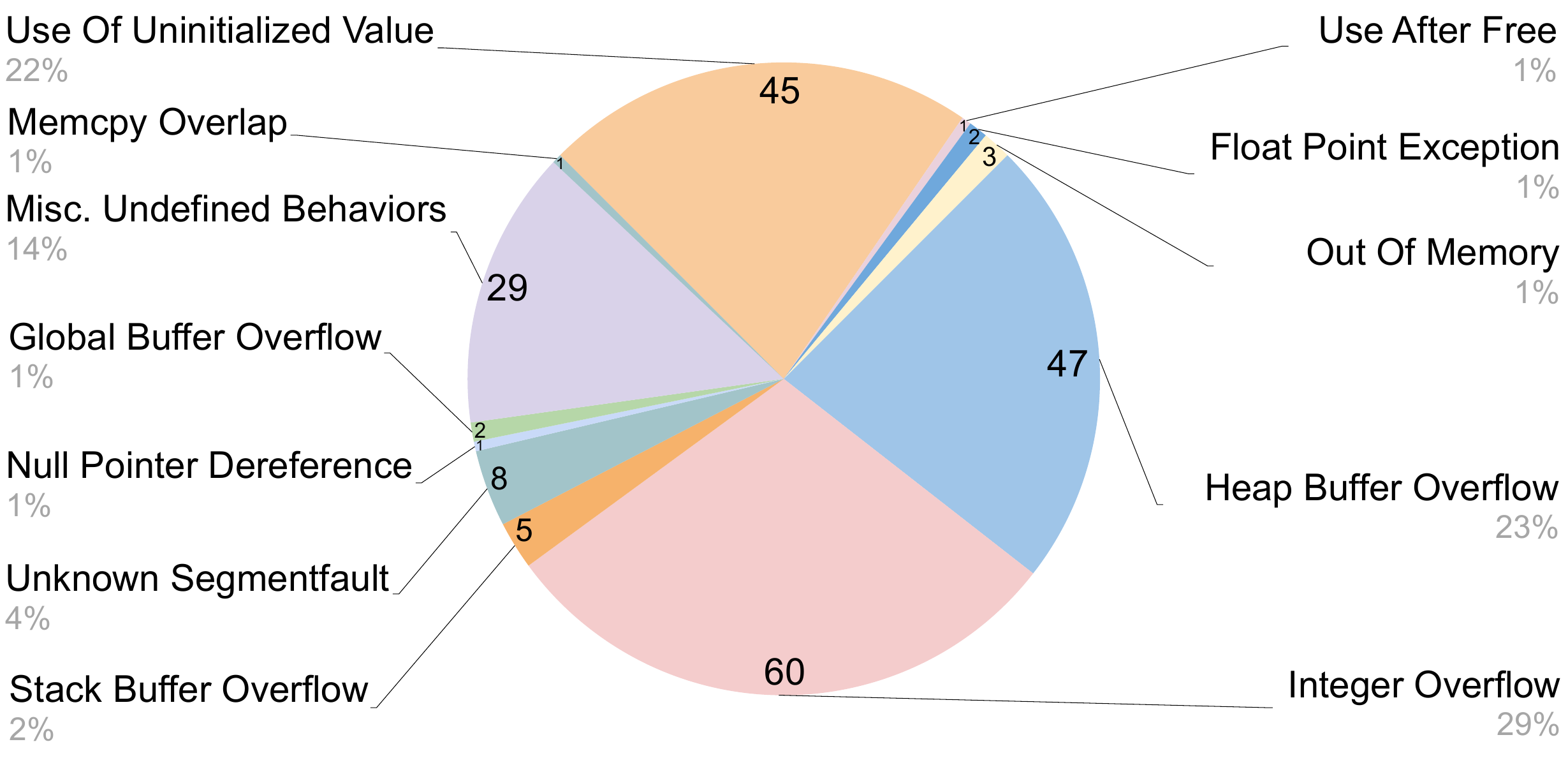}
        \vspace{-5pt}
        \caption{\revise{Distribution of bug types found by \toolname.}}
        \vspace{-5pt}
        \label{fig:bugtype}
    \end{minipage}
\end{minipage}
\vspace{-10pt}

\end{figure*}

\smallskip
\para{Number of Unique Bugs.} 
We plot the number of unique bugs in every repetition in Figure~\ref{fig:box-bugs}.
Table~\ref{tab:bugs-mean} aggregates the results and reports the mean number of unique bugs.
\revise{
On 13 out of 20 programs, \toolname found more bugs than all other fuzzers. For the remaining 7 programs, there is no statistical difference between \toolname and the second-best fuzzer.
Since \fuzzpost is the overall second-best fuzzer, we report the Mann-Whitney U-test~\cite{utest} ($p$-val) and Vargha-Delaney~\cite{a12} effect size (\vdtest) when comparing \toolname with it.
The $p$-val column shows that \toolname is statistically different from \fuzzpost ($p$-val < 0.05) on 12 programs. There is only one case (\mpgain) where \toolname has a lower mean number of bugs ($7.75$ v.s $7.80$). However, there is no statistical significance ($p$-val=$0.72$) between them.
Intuitively, the \vdtest column measures to what extent or the probability that \toolname is better than \fuzzpost. The result shows that 8 programs have \vdtest > $0.71$ (conventionally large effect size) and 5 programs have \vdtest > $0.64$ (conventionally medium effect size). Notably, there is no program with \vdtest < $0.44$, meaning that \fuzzpost cannot surpass \toolname on any programs. 
}
In summary, our result answers \ref{rq:bug-num}: \emph{\toolname has a significantly stronger bug-finding capability than all other fuzzers.} 


\smallskip
\para{Accumulative Number of Unique Bugs.} 
To understand the overlaps of bugs found by different fuzzers, we accumulate all unique bugs found by each fuzzer in each repetition.
\revise{
Table~\ref{fig:bugs-venn} shows the unique number of bugs found by each fuzzer. It shows that \toolname \emph{covers the most number of bugs (204)}. 
The ``$\toolname-$'' shows that \textit{\toolname does not miss any bugs found by any other fuzzers.}
The ``$\toolname+$'' shows that \textit{\toolname always finds more bugs than other fuzzers.} Compared to the second-best fuzzer, \fuzzpost, \toolname found 30 more bugs, which also confirms that simply post-processing coverage-increasing inputs will miss many bugs.}
In summary, we can answer \ref{rq:bug-num} and \ref{rq:bug-miss}: \emph{\toolname does not miss any bugs and can find significantly more bugs.} We will discuss the potential false negative problem of \toolname later.

\smallskip
\para{Number of Unique Bugs Reported by Sanitizer-enabled Programs.} 
Of all the 204 unique bugs identified by \toolname, more than 65\% are not detectable on normally built programs. These bugs are reported after invoking sanitizer-enabled programs in \toolname, highlighting the necessity of using sanitizers.

\smallskip
\para{Bug Types.}
We now try to understand which types of bugs \toolname can cover. Our observation, as illustrated in Section~\ref{sec:illustrative}, is that all bug types, like Buffer-Overflow and Integer-Overflow, can result from/in execution path changes.
Figure~\ref{fig:bugtype} reports the types of bugs found by \toolname.
The result highlights that \toolname can indeed cover all bug types, such as heap Buffer-Overflow (47 bugs), Integer-Overflow (60 bugs), Use-of-Uninitialized-Memory (45 bugs), and Use-after-Free (2 bugs).
The small number of Use-after-Free bugs is because they are indeed relatively rare in practice~\cite{unifuzz}.

%

\smallskip
\para{False Negatives.} 
\textit{Despite the outstanding performance of \toolname, we have no theoretical guarantee that \toolname can cover all bugs}.
Theoretically, \toolname may have false negatives where certain bugs are missed.
This false negative impact can be inferred from the \mujs performance in Table~\ref{tab:bugs-mean}, where \toolname has a slightly lower mean number of bugs (7.9) than AFL++-\debloat/\ubsan (8.2). 
The reason is that one bug in \mujs was not triggered in every repetition of \toolname, but in every repetition in the other fuzzer.
The execution pattern for this bug can sometimes be seen in normal executions, which leads to a less frequent discovery in \toolname.
Due to the stochastic nature of fuzzing, triggering a bug only once is sufficient for \toolname to detect it in practice.
Given the overall superior performance of \toolname, we believe that the moderate false negative issue is acceptable and does not impede its general effectiveness.

\subsection{Effectiveness of Execution Pattern}\label{sec:eval-pattern}
We now break down \toolname to evaluate the effectiveness of its core design, \ie, execution pattern.
To understand if unique execution patterns can accurately encapsulate bug-triggering inputs/executions, we extensively analyze all executions during fuzzing. Specifically, we conduct the following experiments:
\begin{enumerate}[label={\textbf{Step (\arabic*)}}, labelwidth=!, labelindent=35pt, itemsep=3pt, topsep=3pt]
    \item Use AFL++ to fuzz the \emph{normally built} program.
    \item \label{enum:pattern} For each generated input, we first obtain its execution pattern, then examine whether or not this execution pattern is unique, \ie, has been observed before.
    \item \label{enum:san} For each input, no matter whether or not its execution pattern is unique, 
    \revise{
    we run it on \asan-, \ubsan-, and \msan-enabled programs to test if it triggers a bug.
    }
\end{enumerate}

\revise{
To explore and measure other designs of the execution pattern, we provide the following two alternatives:
\vspace{3pt}
\\~
{\noindent\textbf{Alternative 1: Hit Count}}. Recall our current execution pattern definition in \ref{def:execution-pattern}, where we do not present the hit count of each visited code edge. In this alternative design, we include the hit count of each code edge in the execution pattern. 
\vspace{3pt}
\\~
{\noindent\textbf{Alternative 2: Coverage}}. In this design, the execution pattern is the coverage-increasing information. Whenever an execution increases code coverage, we identify it as sanitizer-required. This design is actually equivalent to the strategy used in \fuzzpost.
}

\revise{
We ran the experiments on our current execution pattern as well as two alternative designs for 24 hours and repeated them twenty times. 
With this set of experiments, for each execution pattern design, we want to answer:
}
\begin{enumerate}[label=\textbf{Q\arabic*}, leftmargin=20pt, topsep=5pt, itemsep=5pt,start=3]
\item\label{rq:num-unique-all} Out of all executions in~\textbf{Step (2)}, how many of them are marked as having unique execution patterns?
\item\label{rq:num-unique-bug} Out of all bug-triggering inputs/executions in~\textbf{Step (3)}, how many of them are also marked as having unique execution patterns in ~\textbf{Step (2)}? 
\end{enumerate} 

The above \ref{rq:num-unique-all} can tell us the ratio of unique execution patterns during fuzzing. A smaller ratio indicates that sanitization is required less frequently, leading to a higher speed.
The following \ref{rq:num-unique-bug} can inform us how effective the unique execution pattern is in filtering bug-triggering inputs.
Table~\ref{tab:bug-pattern} shows the result. 
\revise{
The ``All-\toolname'' column shows that more than half of all programs have less than 1\% unique execution patterns in our current design. The average ratio is only 3.8\%. We can thus conclude that \emph{Only a small fraction of fuzzer-generated inputs have unique execution patterns.}
The ``Bug-\toolname'' column shows that, on average, 91.7\% of bug-triggering inputs have unique execution patterns, which means that if we only pass inputs with unique execution patterns to sanitizer-enabled programs, we can successfully sanitize 91.7\% buggy inputs.
}
{
One abnormal case is \lame with a 42.7\% ratio. The reason is that \lame is not stable, \ie, executing even the same input multiple times can lead to divergent execution paths. In fact, even vanilla AFL++ discourages fuzzing unstable programs~\cite{aflperformance}. 
}
Note that we do not deduplicate all bug-triggering inputs here; most of them are in fact duplicates. 
Considering the same bug can be triggered multiple times during fuzzing, 91.7\% accuracy can already ensure that no bugs will be missed with a high probability.

\revise{
The ``Bug-Hit'' column shows that when we use \textbf{Hit Count} as the execution pattern, averagely 97.1\% of bug-triggering inputs have unique execution patterns (higher than 91.7\% in \toolname), but the average ratio of all inputs (``All-Hit'' column) is as high as 13.1\% (much higher than 3.8\% in \toolname). This means that 13.1\% of inputs will be identified as sanitizer-required, hindering the fuzzing throughput dramatically.
\\\indent
The ``Bug-Cov'' column shows that when we use \textbf{Coverage} as the execution pattern, averagely only 45\% of bug-triggering inputs have unique execution patterns. Such a low ratio will cause many bug-triggering inputs to not be sanitized, consequently missing bugs. 
Our previous bug-finding evaluation of \fuzzpost in Table~\ref{fig:bugs-venn} also shows that using coverage will miss 30 out of 204 bugs. Note that since we do not duplicate all bug-triggering inputs here, 45\% of bug-triggering inputs does not imply 45\% of all bugs.
\\\indent
In summary, compared to the two alternative designs, our current execution pattern design achieves the best balance between the two ratios. 
However, this does not mean that our current design is perfect.
Ideally, the ratio of all inputs should be as low as possible, while the ratio of bug-triggering inputs should remain as high as possible. Exploring better execution pattern design is an exciting future work.
}

\begin{table}[tp]
    \centering
    \small
    \caption{\revise{
    Ratios of inputs that have unique execution patterns. ``All'' refers to the ratio of all generated inputs that have unique execution patterns. ``Bug'' refers to the ratio of bug-triggering inputs that have unique execution patterns.} }
    \renewcommand{\arraystretch}{1.1}
    \rowcolors{4}{gray!15}{}
    \setlength{\tabcolsep}{5pt}
\begin{tabular}[t]{lrrrrrrr}
        \toprule[1.0pt]
         \multirow{2}{4em}{\textbf{Programs}} & \multicolumn{3}{c}{\textbf{All}} && \multicolumn{3}{c}{\textbf{Bug}}\\
         \cline{2-4} \cline{6-8}
         & \textbf{Hit} & \textbf{Cov} & \textbf{\toolname} && \textbf{Hit} & \textbf{Cov} & \textbf{\toolname} \\
         \hline
cflow & $ 17.7 $ & < $ 0.1 $ & $ 2.3 $ && $ 100.0 $ & $ 2.8 $ & $ 100.0 $ \\
exiv2 & $ 3.2 $ & < $ 0.1 $ & $ 0.2 $ && $ 87.4 $ & $ 23.2 $ & $ 75.3 $ \\
ffmpeg & $ 11.3 $ & < $ 0.1 $ & $ 0.8 $ && $ 100.0 $ & $ 75.2 $ & $ 99.9 $ \\
gdk-pixbuf. & $ 2.0 $ & < $ 0.1 $ & < $ 0.1 $ && $ 89.2 $ & $ 65.6 $ & $ 86.7 $ \\
imginfo & $ 0.6 $ & < $ 0.1 $ & < $ 0.1 $ && $ 82.8 $ & $ 29.9 $ & $ 60.0 $ \\
infotocap & $ 20.7 $ & < $ 0.1 $ & $ 6.5 $ && $ 99.9 $ & $ 7.7 $ & $ 95.3 $ \\
jhead & $ 8.6 $ & < $ 0.1 $ & $ 1.0 $ && $ 100.0 $ & $ 13.1 $ & $ 92.3 $ \\
jq & $ 10.1 $ & < $ 0.1 $ & $ 3.1 $ && $ 98.3 $ & $ 0.6 $ & $ 60.6 $ \\
sqlite3 & $ 25.4 $ & < $ 0.1 $ & $ 7.5 $ && $ 100.0 $ & $ 69.3 $ & $ 100.0 $ \\
lame & $ 72.2 $ & < $ 0.1 $ & $ 42.7 $ && $ 88.5 $ & $ 80.8 $ & $ 88.5 $ \\
mp3gain & $ 40.7 $ & < $ 0.1 $ & $ 0.4 $ && $ 99.3 $ & $ 64.2 $ & $ 89.8 $ \\
mp42aac & < $ 0.1 $ & < $ 0.1 $ & < $ 0.1 $ && $ 100.0 $ & $ 52.6 $ & $ 100.0 $ \\
mujs & $ 20.1 $ & < $ 0.1 $ & $ 4.7 $ && $ 100.0 $ & $ 25.9 $ & $ 98.7 $ \\
nm & $ 1.0 $ & < $ 0.1 $ & $ 0.2 $ && $ 100.0 $ & $ 59.5 $ & $ 94.4 $ \\
flvmeta & $ 0.3 $ & < $ 0.1 $ & < $ 0.1 $ && $ 98.6 $ & $ 2.8 $ & $ 97.2 $ \\
objdump & $ 4.2 $ & < $ 0.1 $ & $ 0.7 $ && $ 99.9 $ & $ 83.6 $ & $ 99.9 $ \\
pdftotext & $ 20.2 $ & < $ 0.1 $ & $ 3.7 $ && $ 100.0 $ & $ 84.6 $ & $ 99.9 $ \\
tcpdump & $ 2.8 $ & < $ 0.1 $ & $ 0.8 $ && $ 99.8 $ & $ 70.9 $ & $ 99.7 $ \\
tiffsplit & $ 1.1 $ & < $ 0.1 $ & $ 0.5 $ && $ 99.0 $ & $ 18.0 $ & $ 98.5 $ \\
wav2swf & < $ 0.1 $ & < $ 0.1 $ & < $ 0.1 $ && $ 100.0 $ & $ 70.1 $ & $ 97.7 $ \\
\hline
\hiderowcolors
\textbf{Average} & $13.1$ & $0.02$ & $3.8$ && $97.1$ & $45.0$ & $91.7$ \\
\bottomrule

    \end{tabular}
\label{tab:bug-pattern}
\vspace{-15pt}
\end{table}

\begin{figure*}[tp]
    \centering
    \includegraphics[trim={0 30 0 40}, width=\textwidth, page=1]{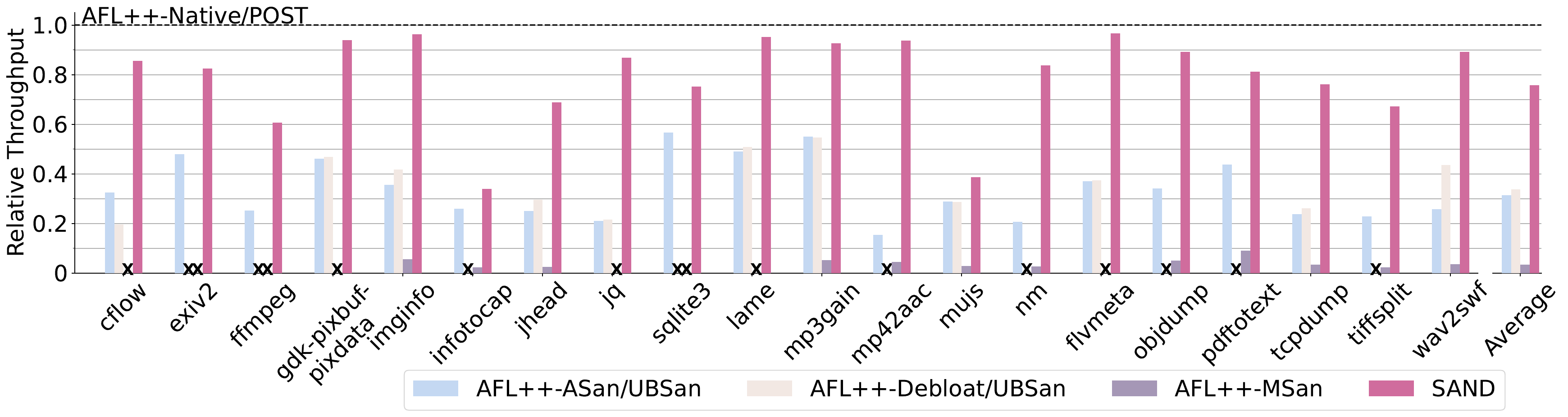}
    \caption{\revise{
    Relative throughput normalized to AFL++-Native/POST. \xmark indicates a compilation failure or an incompatible sanitizer. "Average" refers to the average throughput of all programs.}}
    \label{fig:throughput}
\vspace{-10pt}
\end{figure*}

\subsection{Fuzzing Throughput}\label{sec:throughput}
We analyzed the end-to-end fuzzing throughput, \ie, the total number of inputs executed during fuzzing.
Figure~\ref{fig:throughput} shows the average throughput normalized to AFL++-Native. Since \fuzzpost has the same throughput as AFL++-Native, we use AFL++-Native/POST to represent both.

\smallskip
\noindent\textbf{Compared to Sanitizers-enabled Fuzzers.}
\toolname achieves an average of 2.4x, 2.1x, and 20.0x throughput than AFL++-\asan/\ubsan, AFL++-\debloat/\ubsan, and AFL++-\msan, respectively. Moreover, \toolname has significantly higher throughput \emph{on all programs}.
On some programs, the speedup rate is even higher. For instance, on \nm, \toolname executes 4.2x and 31.8x more inputs than AFL++-\asan/\ubsan and AFL++--\msan, respectively.
\emph{It is worth mentioning that \toolname is equipped with all three sanitizers, including the slowest \msan.}

\smallskip
\noindent\textbf{Compared to AFL++-Native/POST.} 
Overall, \toolname achieves 75\% of AFL++-Native's throughput. On 6 programs, \toolname achieves more than 90\% of AFL++-Native's throughput. 
\emph{The result shows that \toolname successfully increases the speed of fuzzing on sanitizer-enabled programs to a near-native level.}

\subsection{Code Coverage}\label{sec:coverage}

We use the {\texttt{afl-showmap}} utility in AFL++ to collect the code coverage. Table~\ref{tab:cov} presents the average code coverage achieved by each fuzzer on each program.

\smallskip
\noindent\textbf{Compared to Sanitizers-enabled Fuzzers.} 
\toolname achieves higher coverage on \emph{all programs}.
Intuitively, since \toolname has a much higher throughput than other sanitizer-enabled fuzzers, \toolname executes more inputs and thus achieves higher coverage.

\smallskip
\noindent\textbf{Compared to AFL++-Native/POST.} 
On 11 out of 20 programs, there is no significant coverage difference between AFL++-Native/POST and \toolname. For the remaining 9 programs, \toolname achieves almost the same code coverage as AFL++-Native. 
As analyzed before, \toolname can achieve 75\% throughput of AFL++-Native, which accounts for the coverage drop in some programs. 
Since bug-finding capability is the golden metric for fuzzing, although AFL++-Native/POST can achieve higher code coverage, it has a worse bug-finding rate and thus is less favorable in practice.


\begin{table}[tp]
    \centering
    \small
    \renewcommand{\arraystretch}{1.0}
    \rowcolors{4}{gray!15}{}
    \caption{\revise{
    Code coverage (\%) of fuzzers with statistical $p$-val. \xmark indicates a compilation failure. The highest code coverage compared to AFL++-Native is highlighted in \green{green}.}}
    \vspace{-5pt}
    \setlength\tabcolsep{3pt}
    \begin{tabular}{lr|llll}
        \toprule[1.0pt]
        \multirow{2}{*}{\textbf{Programs}} & \multirow{2}{*}{\makecell[l]{\textbf{AFL++}\\\textbf{-Native}\\/\textbf{-POST}}} & \multicolumn{3}{c}{\textbf{AFL++-}} & \multirow{2}{*}[-0.2em]{\textbf{\toolname}} \\
        \cline{3-5}
         & & \makecell[l]{\textbf{ASan/}\\\textbf{UBSan}} & \makecell[l]{\textbf{Debloat/}\\\textbf{UBSan}} & \textbf{\msan} & \\
        \hline
cflow & 24.30 & $23.93_{0.00}$ & $21.95_{0.00}$ & \xmark & \green{24.26}$_{0.10}$ \\
exiv2 & 6.02 & $3.13_{0.00}$ & \xmark & \xmark & \green{5.83}$_{0.16}$ \\
ffmpeg & 2.70 & $0.31_{0.00}$ & \xmark & \xmark & \green{2.61}$_{0.00}$ \\
gdk-pixbuf. & 21.49 & $18.54_{0.00}$ & $18.13_{0.00}$ & \xmark & \green{21.32}$_{0.67}$ \\
imginfo & 13.17 & $11.53_{0.00}$ & $11.82_{0.00}$ & $8.61_{0.00}$ & \green{12.75}$_{0.16}$ \\
infotocap & 20.33 & $11.40_{0.00}$ & \xmark & $13.55_{0.00}$ & \green{18.87}$_{0.01}$ \\
jhead & 15.31 & $9.96_{0.00}$ & $9.68_{0.00}$ & $15.18_{0.00}$ & \green{15.31}$_{1.00}$ \\
jq & 31.11 & $30.00_{0.00}$ & $28.50_{0.00}$ & \xmark & \green{31.14}$_{0.86}$ \\
sqlite3 & 43.34 & $36.81_{0.00}$ & \xmark & \xmark & \green{38.93}$_{0.00}$ \\
lame & 31.96 & $31.29_{0.00}$ & $28.31_{0.00}$ & \xmark & \green{31.88}$_{0.24}$ \\
mp3gain & 41.31 & $39.40_{0.00}$ & $39.14_{0.00}$ & $34.34_{0.00}$ & \green{40.38}$_{0.01}$ \\
mp42aac & 7.09 & $5.11_{0.00}$ & \xmark & $7.50_{0.00}$ & \green{7.01}$_{0.13}$ \\
mujs & 28.18 & $16.08_{0.00}$ & $16.51_{0.00}$ & $20.35_{0.00}$ & \green{27.00}$_{0.00}$ \\
nm & 7.82 & $3.48_{0.00}$ & \xmark & $6.32_{0.00}$ & \green{7.58}$_{0.00}$ \\
flvmeta & 5.28 & $3.36_{0.00}$ & $3.19_{0.00}$ & \xmark & \green{5.28}$_{1.00}$ \\
objdump & 6.96 & $6.40_{0.00}$ & \xmark & $6.40_{0.00}$ & \green{6.87}$_{0.00}$ \\
pdftotext & 16.46 & $13.18_{0.00}$ & \xmark & $14.92_{0.00}$ & \green{15.34}$_{0.00}$ \\
tcpdump & 18.76 & $9.34_{0.00}$ & $8.65_{0.00}$ & $11.74_{0.00}$ & \green{15.94}$_{0.01}$ \\
tiffsplit & 21.06 & $18.25_{0.00}$ & \xmark & $12.31_{0.00}$ & \green{20.91}$_{0.33}$ \\
wav2swf & 1.92 & $1.91_{0.00}$ & $1.87_{0.00}$ & $1.61_{0.00}$ & \green{1.89}$_{0.32}$ \\
\bottomrule
    \end{tabular}
    \label{tab:cov}
\vspace{-10pt}

\end{table}


\subsection{\revise{Hash Function}}\label{sec:hash-overhead}
\revise{
We evaluated both the \textit{hash overhead} and \textit{hash collision} to demonstrate that our selected hash function is sufficiently effective to support \toolname. 
Our first evaluation measured the overall time spent on the hash function during fuzzing. The result shows that \emph{hashing operations in \toolname have an average of 1\% overhead, which is negligible in fuzzing}.
Our second evaluation measured if there was any hash collision, \ie, two different execution patterns have the same hash value. The result shows that nearly \emph{no hash collision was detected during 24 hours of fuzzing}.
These evaluations convincingly demonstrate the effectiveness of our selected hash function.
}






\section{Discussion}\label{sec:discussion}

\noindent\textbf{Compatibility to Other Advanced Fuzzers.}
\toolname does not touch the main fuzzing logic of a CGF. It is orthogonal to many other fuzzer advances. For example, new mutation strategies~\cite{aschermann2019redqueen,mopt}, effective seed scheduling schemes~\cite{aflfast,bohme2020boosting}, and hybrid fuzzing techniques~\cite{huang2020pangolin,stephens2016driller} can all be normally integrated into a CGF fuzzer, which \toolname can build upon.
\revise{
To understand \toolname’s general applicability, we port it to an alternative fuzzer MOpt~\cite{mopt}, which uses a different mutation scheduling strategy and can be manually turned on in AFL++. We include the details in the supplementary.
}


\para{Incompatibility to Coverage-guided Tracing.}
Our current execution pattern is collected from the coverage bitmap. Some research efforts are trying to reduce coverage collection overhead, such as HexCite~\cite{hexcite} and UnTracer~\cite{untracer}. Such approaches break the coverage map and thus cannot be used together with \toolname.
However, we would like to highlight that the coverage collection overhead is much smaller compared to sanitizers. 
Researchers~\cite{odin} have shown that the latest coverage collection approach used in AFL++ only brings a median of 15\% overhead. Sanitizers like \asan and \msan can incur 237$\sim$6,836\% overhead.
Even if these approaches can entirely eliminate coverage tracing overhead, the overall benefit when sanitizers are used is small.

\para{Limitations.}
Despite that \toolname brings significant improvement to fuzzing, it also comes with a few limitations.
The first limitation is the gap between the unique execution pattern ratio and the bug-triggering input ratio. 
Our empirical evaluation in Section~\ref{sec:rare} has shown that many bug-triggering input ratios are below 0.5\%, which is lower than the average unique execution pattern ratio of 3.8\%. This gap indicates that there is still space for improvement. Designing more effective execution abstraction is an interesting future work.
The second limitation is that although our evaluation has confirmed that \toolname did not miss any bugs, we can not provide a theoretical guarantee. It would be interesting and useful to explore sound execution analysis to eliminate this concern.

\section{Related work}

\para{Reducing Sanitizer Overhead.} 
ASAP~\cite{wagner2015high} removes sanitizer checks to meet a required performance budget.
FuZZan~\cite{fuzzan} dynamically selects an optimal metadata structure for \asan and \msan.
SanRazor~\cite{sanrazor} and Debloat~\cite{debloat} remove redundant sanitizer checks via either static or dynamic analysis. SanRazor supports both \asan and \ubsan while \debloat only supports \asan.
All of these techniques require significant modifications to sanitizer implementations, which inevitably hinders their practical adoption.
\toolname, on the other hand, uses sanitizers without any modification, which further highlights the orthogonality of \toolname to these efforts. 

\para{Bug pattern.}
Igor~\cite{igor} observes that all bug-triggering inputs have unusual execution behaviors. For specific bug types, UAFL~\cite{uaf} prioritizes memory operations of longer sequences to effectively detect User-After-Free bugs. Dowser~\cite{dowser} selectively checks instructions that access arrays in a loop for detecting buffer overflow bugs. ParmeSan~\cite{parmesan} leverages the knowledge from sanitizer instrumentations to discover certain types of bugs faster. PGE~\cite{pge} finds that bug-triggering executions correlate with execution prefixes.
At a high level, the findings or insights behind these approaches share similar motivations to our execution pattern, \ie, bug-triggering inputs tend to have unique execution features.

\para{Improving Fuzzing Performance.}
Since the success of AFL~\cite{zalewski2014afl}, the fuzzing community has seen a broad range of new fuzzer developments. In particular, coverage-guided grey-box fuzzers such as AFL++~\cite{aflpp}, AFLFast~\cite{aflfast}, and AFLGo~\cite{aflgo} are the most widely adopted and studied fuzzing techniques. 
Researchers have also put great efforts into optimizing various aspects of fuzzing, such as seed scheduling~\cite{bohme2020boosting}, mutation strategies~\cite{mopt, aschermann2019redqueen}, and path explorations~\cite{stephens2016driller, huang2020pangolin}.
In theory, all these improvements are not related to sanitizer-enabled programs and, therefore, are orthogonal to us.

\para{Reducing Coverage Collection Overhead.} 
Some researchers point out that coverage collection in fuzzing brings extra overhead.
Untracer~\cite{untracer} and HexCite~\cite{hexcite} remove instrumentation code in visited code edges to reduce coverage collection overhead. Zeror~\cite{zeror} shifts between diversely-instrumented binaries to achieve low coverage collection overhead on most executions.
Odin~\cite{odin} dynamically recompiles a binary when the instrumentation requirement changes.
Because all these approaches need to modify the coverage bitmap, our approach is not compatible with them. 
As discussed in Section~\ref{sec:discussion}, coverage collection cost is rather small compared to sanitizer overhead, and thus, our approach is more beneficial.

\section{Conclusion}
We have presented a new fuzzing framework, \toolname, to decouple sanitization from the fuzzing loop. 
\toolname performs fuzzing on the normally built program and only executes sanitizer-enabled programs when an input is identified as sanitization-required.
\toolname utilizes the fact that most of the fuzzer-generated inputs do not need sanitization, which enables it to spend most of the fuzzing time on normally built programs.
We have evaluated \toolname on 20 real-world programs to demonstrate its superiority.
Our work represents an exciting research direction toward the overhead-free adoption of sanitizers in fuzzing.

\bibliographystyle{plain}
\bibliography{ref}

\end{document}



\title{
Supplementary for \\
{\LARGE ``\toolname: Decoupling Sanitization from Fuzzing for Low Overhead''}
}




\maketitle

\section{Hash in \toolname}\label{sec:hash-overhead}

As Algorithm~1 indicates, \toolname utilizes a hash table to store hash checksums for each execution pattern.
In this section, we evaluate the hash overhead of \toolname and the potential hash collision risk in the hash table.

\medskip
\para{Hash Overhead.}
We modified \toolname to two versions to precisely evaluate the hash overhead. First, \textsc{NoHash}, where lines 6-11 in Algorithm~1 are removed so that no hash operations are performed. Second, \toolname-\textsc{Hash}, where lines 8-11 in Algorithm~1 are removed so that hash operations are performed as the normal \toolname, but no sanitizer-enabled programs are invoked.
We use the same random seed for both modified fuzzers to generate an identical set of inputs on the same initial seed pool.
We run both fuzzers on each program ten times and record the total fuzzing time on the first \emph{one million} inputs.
The second and third columns in Table~\ref{tab:hashing} report the average speed of both fuzzers. On 14 out of 20 programs, both fuzzers do not have statistically differentiable ($p$-val >= 0.05) speed. Only on \cflow~, \ffmpeg, and \sqlite, \toolname-\textsc{Hash} is slightly slower at a rate of 3.13\%m 3.69\%, and 2.58\% while averagely \toolname-\textsc{Hash} only incurs 1.11\% penalty.
We can then conclude that \emph{hashing operations in \toolname have negligible overhead to fuzzing.}

\medskip
\para{Hash Collision.}
\emph{Hash collision} can happen when two different execution patterns either have the same hash checksum or result in the same index in the hash table.
Because the effectiveness of \toolname relies on the accurate identification of unique execution patterns, hash collisions may potentially harm the performance.
To evaluate the hash collision rate of \toolname, we use the \toolname-\textsc{Hash} and expand it to save all execution patterns (rather than checksums) to disk. When an execution pattern is marked as observed, we compare this execution pattern with the saved execution pattern byte to byte. Any difference in the comparison signifies a hash collision.
The last column in Table~\ref{tab:hashing} lists the number of hash collisions for the first \emph{one million} inputs. The result shows that \emph{no hash collision} was detected on 18 programs. The main reason is that unique execution patterns are rare (averagely 3.1\% according to Section~II-B), making the hash table sparse and hard to have collisions. On \sqlite and \lame, only 8.5 and 21 hash collisions were detected. Consider the often millions of or even more hash operations during fuzzing, dozens of collisions do not have an observable effect.

\begin{figure}[htp]
\centering
    \subfloat[Average throughput.\label{fig:mopt-throughput}]{
        \includegraphics[width=0.20\textwidth]{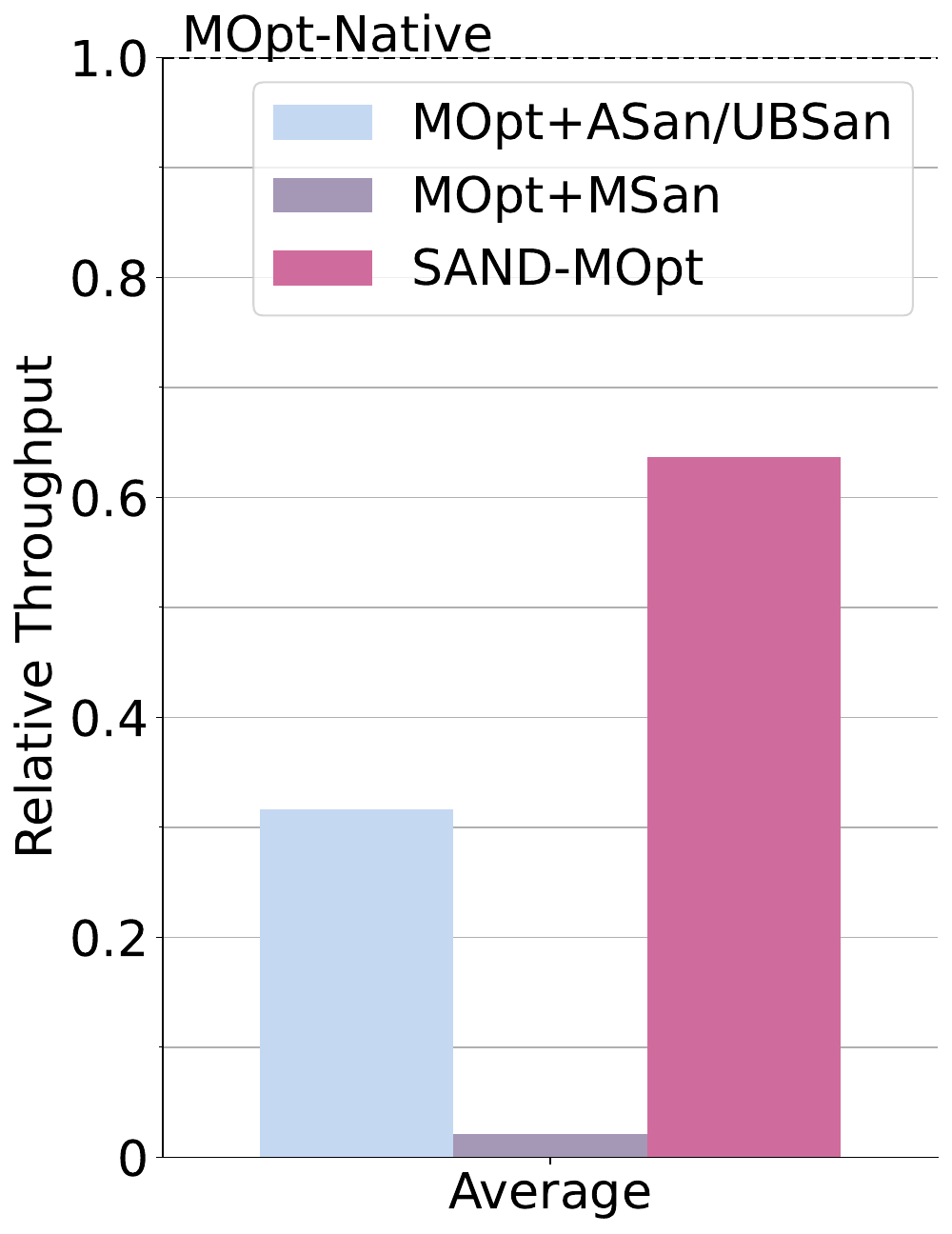}
    }
    \hfill
    \subfloat[Unique bugs breakdown.\label{fig:mopt-bugs}]{
        \includegraphics[trim=60 0 80 0, width=0.20\textwidth, page=7]{imgs/Figures_old.pdf}
    }
    \caption{The overall performance of \toolname-MOpt.}
    \label{fig:mmopt-figs}
\end{figure}

\begin{table}[tp]
    \centering
    \small
    \caption{The hash overhead and collision in \toolname.}
    \rowcolors{4}{gray!15}{}
    \renewcommand{\arraystretch}{1.2}
    \begin{tabular}{lrrrc}
        \toprule[1.0pt]
         \multirow{2}{*}[-0.5em]{\textbf{Programs}} & \textbf{\textsc{NoHash}} & \multicolumn{3}{c}{\textbf{\toolname-\textsc{Hash}}}  \\
         \cline{2-2}  \cline{3-5}
         & \textbf{Speed} & \textbf{Speed}\textsubscript{$p$-val} & \textbf{Overhead} & \textbf{Collision} \\
         \hline
cflow & 3,945 & 3,826\textsubscript{\ 0.00} & 3.13\%  & 0 \\ 
exiv2 & 2,103 & 2,064\textsubscript{\ 0.07} & 1.84\%  & 0 \\ 
ffmpeg & 318 & 307\textsubscript{\ 0.00} & 3.69\%  & 0 \\ 
gdk-pixbuf-pixdata & 681 & 681\textsubscript{\ 0.94} & 0.00\%  & 0 \\ 
imginfo & 3,114 & 3,100\textsubscript{\ 0.65} & 0.48\%  & 0 \\ 
infotocap & 3,011 & 2,982\textsubscript{\ 0.15} & 0.98\%  & 0 \\ 
jhead & 3,327 & 3,327\textsubscript{\ 0.94} & 0.04\%  & 0 \\ 
jq & 359 & 359\textsubscript{\ 0.00} & 0.21\%  & 0 \\ 
sqlite3 & 2,421 & 2,361\textsubscript{\ 0.05} & 2.58\%  & 8.5 \\ 
lame & 210 & 208\textsubscript{\ 0.00} & 0.95\%  & 21 \\ 
mp3gain & 1,929 & 1,915\textsubscript{\ 0.43} & 0.78\%  & 0 \\ 
mp42aac & 1,702 & 1,688\textsubscript{\ 0.21} & 0.89\%  & 0 \\ 
mujs & 1,841 & 1,812\textsubscript{\ 0.01} & 1.58\%  & 0 \\ 
nm & 2,421 & 2,389\textsubscript{\ 0.26} & 1.37\%  & 0 \\ 
flvmeta & 3,826 & 3,804\textsubscript{\ 0.05} & 0.66\%  & 0 \\ 
objdump & 1,279 & 1,266\textsubscript{\ 0.36} & 1.09\%  & 0 \\ 
pdftotext & 408 & 405\textsubscript{\ 0.00} & 0.66\%  & 0 \\ 
tcpdump & 2,018 & 2,004\textsubscript{\ 0.52} & 0.76\%  & 0 \\ 
tiffsplit & 2,335 & 2,334\textsubscript{\ 0.94} & 0.08\%  & 0 \\ 
wav2swf & 2,830 & 2,817\textsubscript{\ 0.36} & 0.52\%  & 0 \\ 
\hline
\textbf{Average} & 2,004 & 1,982 & \textbf{1.11\%} & \textbf{0} \\ 
\bottomrule
    \end{tabular}
    \label{tab:hashing}
\end{table}

\section{General Applicability}

As depicted in Figure~5, \toolname only augments the baseline CGF fuzzer with a conditional sanitization step. Since no other part of the fuzzer is modified, \toolname is, in principle, generally applicable to other CGF or AFL-family fuzzers.
At a high level, in the sequence of all mutated inputs during fuzzing, \toolname's effectiveness depends on the fact that bug-triggering inputs are likely to have unique execution patterns.
Therefore, different mutation strategies may affect \toolname's performance.
To understand \toolname's general applicability, we port it to an alternative fuzzer MOpt, which uses a different mutation scheduling strategy and can be manually turned on in AFL++. 
We denote this new fuzzer as \toolname-MOpt.
We conduct the evaluation on the same 20 programs and compare them against MOpt-Native, MOpt-\asan/\ubsan, and MOpt-\msan.

Figure~\ref{fig:mopt-throughput} shows the normalized throughput of each fuzzer averaged across all programs.
The results show that \toolname-MOpt has achieved 2.2x and 16x higher throughput than MOpt-\asan/\ubsan and MOpt-\msan, respectively. 
We also count the number of unique bugs found by each fuzzer in Figure~\ref{fig:mopt-bugs}. With no surprise, \toolname-MOpt still discovers many more bugs than any other fuzzers. In particular, \toolname-MOpt covers \textbf{36} extra bugs that are not found by any other fuzzers.
These extraordinary results highlight the effectiveness and general applicability of \toolname.

